\documentclass[twocolumn,final,prb,showpacs,superscriptaddress,floatfix,amsmath,amssymb,amsfonts]{revtex4-1}

\usepackage{graphicx}
\usepackage{multirow}

\usepackage{epstopdf}
\usepackage{epsfig}
\usepackage{array}
\usepackage{verbatim}

\usepackage{amsmath}
\usepackage{amssymb}
\usepackage{amsfonts}
\usepackage{latexsym}

\usepackage{booktabs}

\usepackage{color}

\usepackage[colorlinks,bookmarks=false,citecolor=blue,linkcolor=red,urlcolor=blue]{hyperref}
\usepackage{verbatim}

\def\be{\begin{equation}}
\def\ee{\end{equation}}
\def\bea{\begin{eqnarray}}
\def\eea{\end{eqnarray}}
\def\bs{\boldsymbol}
\def\vec{\mathbf}
\def\mc{\mathcal}

\begin{document}
\title{Magnetoelectric effects in single crystals of the cubic ferrimagnetic helimagnet Cu$_2$OSeO$_3$}

\author{M. Belesi}\email{m.e.belesi@ifw-dresden.de}
\affiliation{Institut de Physique de la Mati\`{e}re Condens\'{e}e, Ecole Polytechnique F\'{e}d\'{e}rale de Lausanne, Station 3, CH-1015 Lausanne-EPFL, Switzerland}
\affiliation{Leibniz Institute for Solid State and Materials Research, Helmholtzstrasse 20, 01069 Dresden, Germany}
\author{I. Rousochatzakis}\affiliation{Leibniz Institute for Solid State and Materials Research, Helmholtzstrasse 20, 01069 Dresden, Germany}
\author{M. Abid}\affiliation{Institut de Physique de la Mati\`{e}re Condens\'{e}e, Ecole Polytechnique F\'{e}d\'{e}rale de Lausanne, Station 3, CH-1015 Lausanne-EPFL, Switzerland}
\author{ U. K. R\"o{\ss}ler}\affiliation{Leibniz Institute for Solid State and Materials Research, Helmholtzstrasse 20, 01069 Dresden, Germany}
\author{H. Berger}\affiliation{Institut de Physique de la Mati\`{e}re Condens\'{e}e, Ecole Polytechnique F\'{e}d\'{e}rale de Lausanne, Station 3, CH-1015 Lausanne-EPFL, Switzerland}
\author{J.-Ph. Ansermet}\affiliation{Institut de Physique de la Mati\`{e}re Condens\'{e}e, Ecole Polytechnique F\'{e}d\'{e}rale de Lausanne, Station 3, CH-1015 Lausanne-EPFL, Switzerland}

\date{\today}

\begin{abstract}
We present magnetodielectric measurements in single crystals of the cubic spin-1/2 compound Cu$_2$OSeO$_3$.
A magnetic field-induced electric polarization ($\vec{P}$) and a finite magnetocapacitance (MC) is observed at the onset of the magnetically ordered state ($T_c = 59$ K).
Both $\vec{P}$ and MC are explored in considerable detail as a function of temperature (T), applied field $\vec{H}_a$, and relative field orientations
with respect to the crystallographic axes. The magnetodielectric data show a number of anomalies which signal magnetic phase transitions,
and allow to map out the phase diagram of the system in the $H_a$-T plane.
Below the 3up-1down collinear ferrimagnetic phase, we find two additional magnetic phases.
We demonstrate that these are related to the field-driven evolution of a long-period helical phase,
which is stabilized by the chiral Dzyalozinskii-Moriya term $D~ \vec{M} \cdot(\bs{\nabla}\times\vec{M})$ that
is present in this non-centrosymmetric compound.
We also present a phenomenological Landau-Ginzburg theory for the ME$_H$ effect, which is in excellent agreement with experimental data,
and shows three novel features:
(i) the polarization $\vec{P}$ has a uniform as well as a long-wavelength spatial component that is given by the pitch of the magnetic helices,
(ii) the uniform component of $\vec{P}$ points along the vector $(H^yH^z, H^zH^x, H^xH^y)$, and
(iii) its strength is proportional to $\eta_\parallel^2-\eta_\perp^2/2$, where $\eta_\parallel$ is the longitudinal and $\eta_\perp$
is the transverse (and spiraling) component of the magnetic ordering.
Hence, the field dependence of P provides a clear signature of the evolution of a conical helix under a magnetic field.
A similar phenomenological theory is discussed for the MC.
\end{abstract}

\pacs{76.60.Jx,75.50.Gg,75.25.-j,77.84.Bw}

\maketitle

\section{Introduction}
In recent years there has been an extensive experimental and theoretical activity in the field of magnetoelectrics.
The magnetoelectric effect (ME) was originally predicted by Curie in 1894,\cite{Curie}
and describes the induction of electric polarization $\vec{P}$ by a magnetic field and,
vice-versa, the induction of magnetization $\vec{M}$ by an electric field $\vec{E}$.
A phenomenological theory was developed by Landau,\cite{Landau}
who considered the invariants in the expansion of the free energy up to linear terms in the magnetic field.
In this approach, symmetry considerations alone can fix the form of the coupling between $\vec{P}$ and $\vec{M}$.\cite{Katsura,Mostovoy}
Based on such symmetry arguments, Dzyaloshinskii proposed the antiferromagnetic (AFM) compound Cr$_2$O$_3$ as a first candidate for the
observation of the ME effect.\cite{Dzyaloshinskii} Indeed, the electrically induced magnetization (ME$_E$) in Cr$_2$O$_3$
was observed experimentally for the first time by Astrov,\cite{Astrov} and soon after, Rado and Folen\cite{Rado} measured the
the converse effect, i.e  the magnetic-field induced electric polarization (ME$_H$).

Despite the scarcity of compounds with cross-coupled magnetic and electric properties, which is related to the antagonistic requirements
for the existence of both orders,\cite{Wang,Hill,Khomskii} the research in the field has grown enormously.
Since the discovery of ME effects in Cr$_2$O$_3$, more than 100 magnetoelectric compounds have been discovered or synthesized.\cite{Wang}
The motivation behind this activity stems from the novel and fundamental physical phenomena involved, but also for exciting potential
applications, including ME sensors and the electric control of magnetic memories without electric currents
(and thus Joule heating).\cite{Eerenstein,Spaldin,Fiebig,Khomskii,Wang,Gajek,Wood}

At the center of interest have been the so-called multiferroics, which are materials possessing at least two switchable order parameters,
such as electric polarization $\vec{P}$, magnetization $\vec{M}$, and strain.\cite{Smolenskii,Schmid}
Multiferroics show spontaneous ME effects in addition to those induced by external fields.
Despite the coexistence of ferroelectricity and magnetism, very few multiferroic materials exhibit strong coupling
between $\vec{P}$ and $\vec{M}$,\cite{Khomskii} and in most cases the coupling is rather weak.
Typical examples are the perovskites BiMnO$_3$ and BiFeO$_3$,
where the ferroelectric transition temperature is much higher than the magnetic one.\cite{Kimura03prb,Wang03}

A breakthrough in the field came with the discovery of the multiferroics TbMnO$_3$\cite{Kimura03nat} and TbMn$_2$O$_5$,\cite{Hur}
where the ferroelectricity is directly driven by the spin  order. The magnetic state in these compounds is either a spiral (as in TbMnO$_3$,\cite{Kimura03nat} Ni$_3$V$_2$O$_8$,\cite{Lawes05} and MnWO$_4$\cite{Lautenschl,Taniguchi}), or a collinear configuration (as in Ca$_3$(CoMn)$_2$O$_6$\cite{Choi}).
For the understanding of the ME effects in these systems, there has been a number of proposed microscopic mechanisms (such as the spin-current model,\cite{Katsura} the exchange-striction model,\cite{Sergienko,Li} and the electric-current cancelation model\cite{Hu}),
in addition to Landau-Ginzburg phenomenological theories.\cite{Smolenskii,Baryakhtar,Stefanovskii,Mostovoy,Betouras}

Here we focus on the cubic magnetoelectric insulator Cu$_2$OSeO$_3$.\cite{Effenberger,Bos,Larranaga,Gnezdilov,Kobets,Belesi10,Miller,Belesi11,Huang,Maisuradze,Tokura12}
This system crystallizes in the non-centrosymmetric space group \textit{P}2$_1$3,\cite{Effenberger} which allows for piezoelectricity and piezomagnetism
but not for a spontaneous polarization. The structure is a three-dimensional (3D) array of distorted corner-sharing tetrahedra of copper spin $S=1/2$ ions.
The unit cell consists of 16 copper ions which belong to two crystallographically inequivalent groups
denoted here by Cu$_{1}$ and Cu$_{2}$ (Wyckoff positions 4$a$ and 12$\textit{b}$, respectively).
Each tetrahedron comprises three Cu$_2$ and one Cu$_1$ ion.

Polycrystalline samples of this compound were studied by Bos {\it et al.},\cite{Bos} who reported a transition to a ferrimagnetic state at T$_{c}=60$ K.
Based on neutron powder diffraction data, Bos {\it et al.}\cite{Bos} proposed that in this state
all Cu$_2$ moments prefer to align parallel to each other and anti-parallel to the Cu$_1$ moments.
When all Cu spins have their full moment, this 3up-1down state corresponds to a 1/2 magnetization plateau, which is realized
for H$\gtrsim$1 kOe.\cite{Bos} The 3up-1down nature of this plateau was later confirmed by high field  (14 T) $^{77}$Se Nuclear Magnetic Resonance (NMR).\cite{Belesi10}

The presence of a finite ME coupling in Cu$_2$OSeO$_3$ is revealed by an anomaly in the dielectric constant,\cite{Bos,Miller}
at the onset of the magnetically ordered state. At the same time, high resolution synchrotron x-ray powder diffraction data\cite{Bos}
show that the system remains metrically cubic down to 10 K, with no anomalous change in the lattice constant through the magnetic transition.
This is also supported by infrared\cite{Miller} and Raman\cite{Gnezdilov} studies in single crystals.
Furthermore, the NMR study by Belesi {\it et al.}\cite{Belesi10}
demonstrates that there is not any measurable structural distortion even in high magnetic fields.\cite{Belesi10}
These findings strongly suggest that the dielectric constant anomaly is not related to any magnetostructural coupling or polar structural distortions,\cite{Bos}
but is rather driven by the (primary) magnetic order parameter.

The magnetism of Cu$_2$OSeO$_3$ below the 1/2 magnetization plateau is very special.
On the theoretical side, the reason can be understood from its non-centrosymmetric crystal structure
that belongs to Laue class T (23). In cubic magnets from this class, the magnetic (free) energy contains
Lifshitz-type invariants, which impair the homogeneity of the magnetic ordering
and cause a continuous, helical twisting of the magnetic order parameter with a fixed sense of rotation,
as first predicted by Dzyaloshinskii.\cite{Dz64}
Microscopically, the origin of these couplings relies on the weak (relativistic)
Dzyaloshinskii-Moriya (DM) antisymmetric exchange that leads to a slight twisting
of neighboring magnetic moments.\cite{Dz58,Moriya60}

The best known cubic chiral helimagnets are the intermetallic compounds and
alloys with the ``B20'' structure, which belong also to the Laue class T.\cite{Bak80}
In magnetic systems with this crystallographic structure, like MnSi,\cite{Ishikawa76}
(Fe,Co)Si,\cite{Beille83}, or FeGe,\cite{Lebech89}
long-period transverse flat ferromagnetic helices are observed to form the magnetic ground state.
Currently, there is a surge of interest in these systems because complex topological
spin textures composed of topologicical solitons called ``chiral Skyrmions'' have
been predicted to exist in these systems under certain additional
conditions,\cite{Bogdanov89,Bogdanov05,Roessler06} and have been observed in thin films.\cite{Yu10,Yu11}
Furthermore, near the magnetic ordering transitions, a complex sequence of unusual magnetic textures
and properties have been observed,\cite{Wilhelm11,Roessler11,Wilhelm12,Kusaka76,Komatsubara77,Kadowaki82,Ishikawa84,
Gregory92,Lebech95,Thessieu97,Lamago06,Grigoriev06,Grigoriev06a,Muehlbauer09,Neubauer09,Bauer10,
Stishov08,Petrova09,Pappas09,Pappas11,Pfleiderer04,Pedrazzini07,Grigoriev10,Grigoriev11}
which are usually referred to as the ``A-phase''.

Very recently, it has been found that the magnetically ordered ground-state
of Cu$_2$OSeO$_3$ is in fact a long-period helix with a pitch of about 50~nm.\cite{Tokura12}
The magnetism has also been described as very similar to that of the B20 chiral ferromagnets,
and the observation of field-driven skyrmion textures
stabilized in thin film single crystals has been reported by Lorentz-microscopy.\cite{Tokura12}

Theoretically, the chiral long-period modulation is the expected behavior of a magnetic system with a simple
vector order parameter that belongs to one of the crystallographic
classes that allow for the existence of Lifshitz-invariants.
The basic spin-structure is dictated by the isotropic exchange, and can be well described by a collinear
ferrimagnetic order with very large exchange fields of the order of 50-100~T.\cite{Belesi10,Janson12}
In fact, detailed microscopic calculations based on density-functional theory\cite{Janson12}
show that no magnetic frustration should be present in Cu$_2$OSeO$_3$,
although the magnetic multi-sublattice structure appears to allow for geometric frustration.
The weak relativistic DM interactions are a secondary effect that causes
a long-period twisting of this primary ferrimagnetic order.
This is in agreement with the symmetry analysis by Gnezdilov {\it et al.},\cite{Gnezdilov}
who argued that DM interactions must be important to understand the magnetism in Cu$_2$OSeO$_3$.
Therefore, the magnetoelectric insulator Cu$_2$OSeO$_3$ should be understood as
a chiral ferrimagnetic helimagnet, and its magnetization process in a field is that
of the field-driven evolution of a chiral helix.

Here we report an extensive magnetoelectric study on single crystals of Cu$_2$OSeO$_3$.
Our data demonstrate the ME$_H$ effect, as well as a finite magnetocapacitance (MC) which sets in at the onset of the
magnetically ordered state ($T_c = 59$ K). We probe these effects in considerable detail
as a function of temperature (T) and applied field $\vec{H}_a$.
We also study the variation of $\vec{P}$ and MC with respect to the relative orientations of $\vec{H}_a$ with the crystallographic axes,
as well as the relative orientation between magnetic and ac-electric $\vec{E}$ fields (for the MC measurements).

Our ME data manifest a number of anomalies which signal magnetic phase transitions.
From these anomalies we map out the magnetic phase diagram of the system
in the H$_a$-T plane, and demonstrate that there are at least two additional magnetic phases below the 1/2-plateau,
the low-field and the intermediate-field phase.
We show that in the intermediate phase the field dependence of the ME$_H$ data
is fully consistent with the evolution of a chiral conical helix, whose propagation vector is along $\vec{H}_a$.
In the low-field phase, the vanishing of the ME response as H $\to$ 0 suggests a
multi-domain structure of flat transverse helices, where the
propagation vectors of the different helices are pinned along some preferred axes of the system.
The transition between the low-field and the intermediate-field phase can then be ascribed to the complete alignment
of these propagation vectors along $\vec{H}_a$.

The above picture of two additional phases at low fields seems to be consistent up to T$\sim$ 30 K.
At higher temperatures, further anomalies are observed which suggest in particular a complex thermal reorientation of the anisotropy.
We also find a very characteristic double-peak anomaly in the ME$_H$ effect in a very narrow (2K) window close to $T_c$,
which might well be related to the possible presence of the ``A-phase'' mentioned above, which is typically expected close to $T_c$.\cite{Tokura12}

On the theory side, we present a Landau-Ginzburg phenomenological theory which
captures both qualitatively and quantitative features of the ME$_H$ effect, in the intermediate and the 1/2-plateau phase.
In particular, this theory explains the angular dependence of the ME$_H$ effect, as well as the robust sign change
of the polarization P as we go from the intermediate to the 1/2-plateau phase.
In addition, it provides strong evidence that the ME effect in this compound
can be attributed to an exchange striction mechanism which involves the symmetric portion of the anisotropic exchange.
Contrary to expectations, the influence of the DM coupling in the ME effect is heavily diminished  by the long-wavelength nature of the
helimagnetism in Cu$_2$OSeO$_3$.

The theory also predicts that apart from the uniform component of the electric polarization (which is what we measure in the present experiments),
there is also a spatially oscillating component (on a mesoscopic scale)
that is naturally driven by the long-period helimagnetism in Cu$_2$OSeO$_3$.
Finally, a similar Landau-Ginzburg theory is also given for the MC measurements.

Our article is organized as follows. In Sec.~\ref{expdetails} we provide details on our experiments which were performed at EPF-Lausanne.
In Sec. \ref{Magnetization} we present our magnetization measurements.
The quasi-static ME and MC measurements are presented in Secs. \ref{sec:MEH} and \ref{sec:MC}, respectively.
The theoretical interpretation of the ME$_H$ and MC data are given in Secs.~\ref{sec:theoryME} and \ref{sec:theoryMC} respectively.
Finally, a brief summary and discussion of our study is given in Sec.~\ref{sec:disc}.

\section{Experimental details}\label{expdetails}
High quality single crystals of Cu$_2$OSeO$_3$ were grown by the standard chemical vapor phase method. More details about the crystal growth can be found in Ref. [\onlinecite{Belesi10}].
For our measurements we have used two single crystals of Cu$_2$OSeO$_3$,
which we denote in the following by ``Crystal A''  and ``Crystal B''. Both crystals have a thin rectangular plate shape,
with dimensions 1.6$\times$3.6$\times$0.4 mm$^3$ (Crystal A) and 1.6$\times$2.3$\times$0.4 mm$^3$ (Crystal B).
The orientation of the crystal axes with respect to the crystal cuts is determined by single-crystal X-ray diffraction.
In particular, the normal to the widest face of Crystal A is [-554], which is about $6^{\circ}$ off from the body diagonal [-111] axis,
while the normal to the widest face of Crystal B is the [100] axis.

\begin{figure}[t]   
\includegraphics[width=0.48\textwidth]{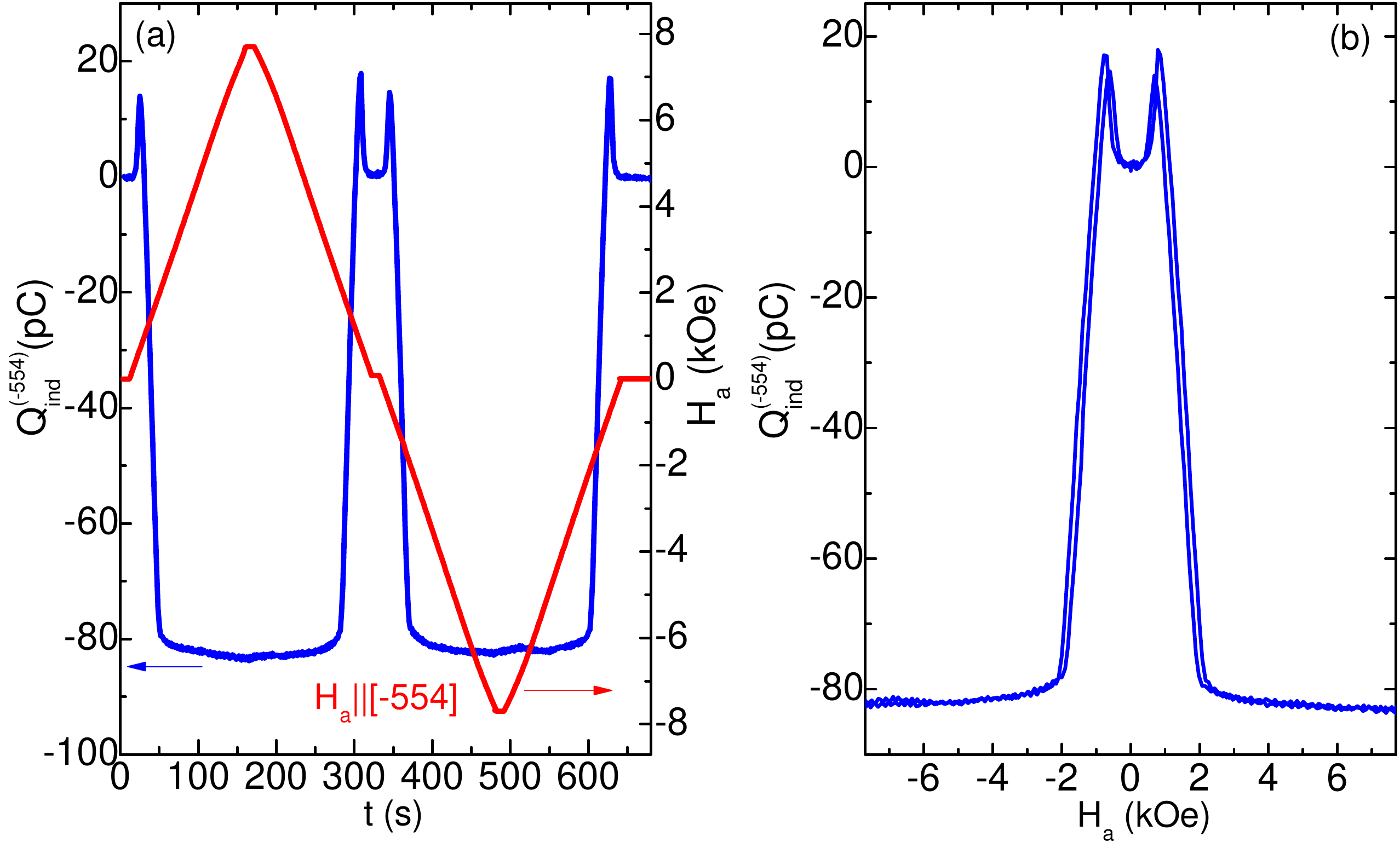}
\caption{(Color online) Demonstration of the magnetoelectric effect in Cu$_2$OSeO$_3$ at 15 K.
(a) The applied magnetic field (right hand axis) is linearly varied with time along the [$-$554] direction,
while the electric charge induced on the ($-$554) surface is measured (left hand axis).
(b) The induced charge from (a) when plotted against the applied field.}
\label{MEdemonstration}
\end{figure}

The magnetization measurements were performed in a Quantum Design Magnetic Property Measurement System (MPMS).
For the dielectric measurements electrical contacts were painted using silver paste.
The capacitance of the crystals was measured using a HP 4284A LCR meter.
For the temperature and magnetic field dependence of the capacitance we have employed a MPMS
and a Janis closed cycle refrigerator, equipped with an electromagnet and a magnetic field regulator.
The latter was also employed for the quasi-static magnetoelectric (ME$_H$) measurements.

For the quasistatic ME$_H$ measurements, a slowly varying magnetic field is applied to the crystal at a fixed temperature and
the induced electric charge Q is recorded with an electrometer. This method is described e.g. by Rivera in Refs. [\onlinecite{Rivera94,Rivera09}].
In a typical experiment, we begin by measuring the charge Q at zero field for a few seconds.
Then, we ramp up the field from 0 to 7.8 kOe at a constant rate (0.08 T/min to 0.4 T/min) while recording the induced charge.
The field is then kept at 7.8 kOe for a few seconds and then the magnetic field is decreased to zero, again linearly with time.
The same procedure can be continued for negative values of the field.
An example of this field cycling is presented in Fig.~\ref{MEdemonstration}(a), which shows an experiment at 15 K with the magnetic field
along the [-554] direction and the induced charge Q measured on the (-554) surface of Crystal A.
Figure \ref{MEdemonstration}(b) shows the induced charge Q as a function of the applied field.
A peak, accompanied by hysteresis, occurs at 700 Oe.
The position of this peak corresponds to the field-induced transition observed in the magnetization process,
and will be discussed in detail in Sec.~\ref{sec:TdependenceME} below.
Finally, the quasistatic ME$_H$ measurements have been performed in the temperature range 10-60 K.

\begin{figure}[t]
\includegraphics[width=0.46\textwidth]{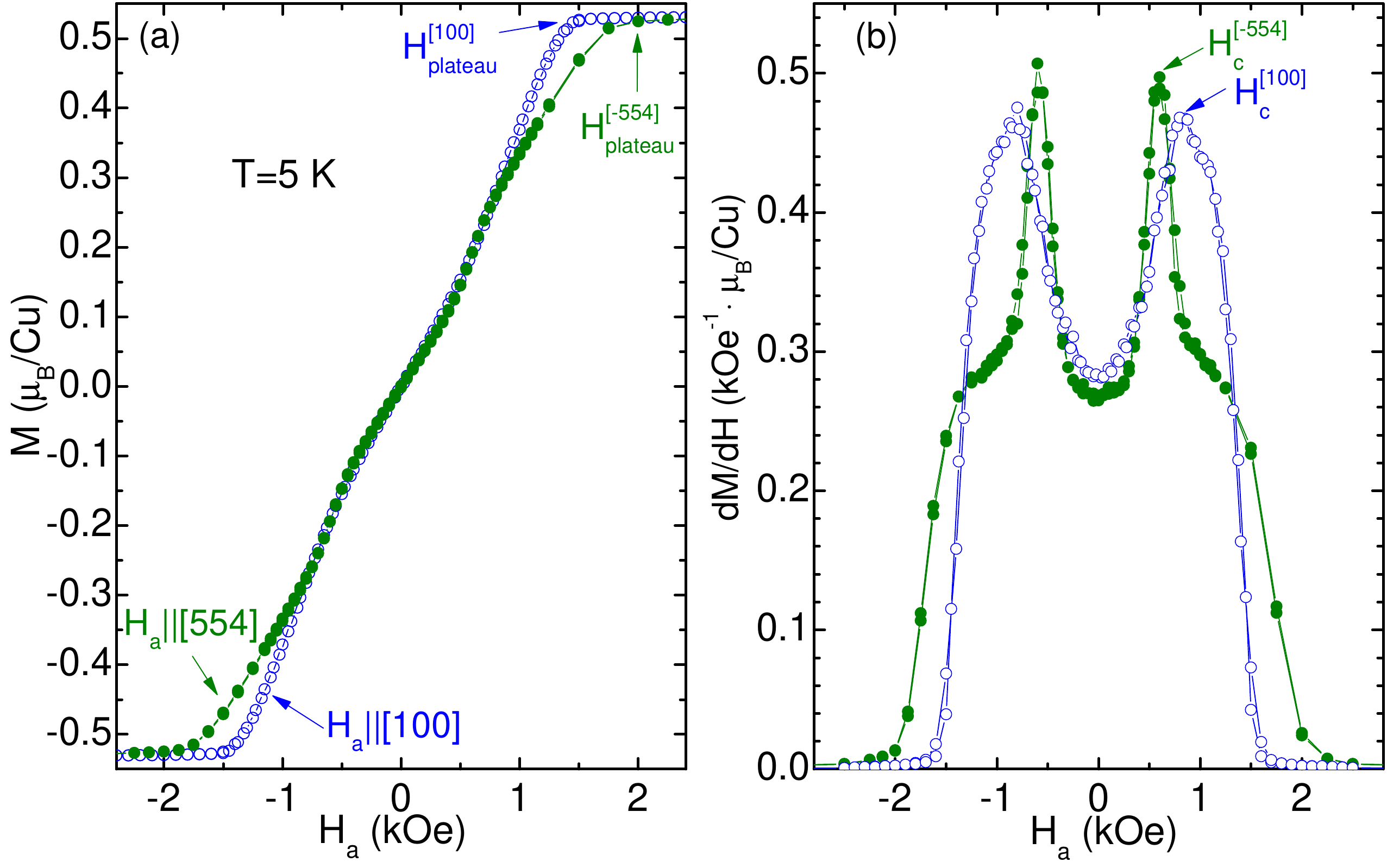}
\caption{(Color online) (a) The field dependence of the magnetization at 5 K in our single crystals. The external magnetic field H$_{a}$ is applied along the [-554] and [100] crystallographic directions, which are perpendicular to the widest rectangular crystal faces.
(b) The first derivatives of the magnetization loops.}
\label{MHloop5K}
\end{figure}

\section{Magnetization measurements}\label{Magnetization}
The magnetic properties of Cu$_2$OSeO$_3$ were first studied by Bos {\it et al.}~\cite{Bos} in polycrystalline samples.
Since then a number of magnetization data in single crystals of Cu$_2$OSeO$_3$ were reported in the literature.\cite{Belesi11,Huang}

\begin{figure}[t]
\includegraphics[width=0.48\textwidth]{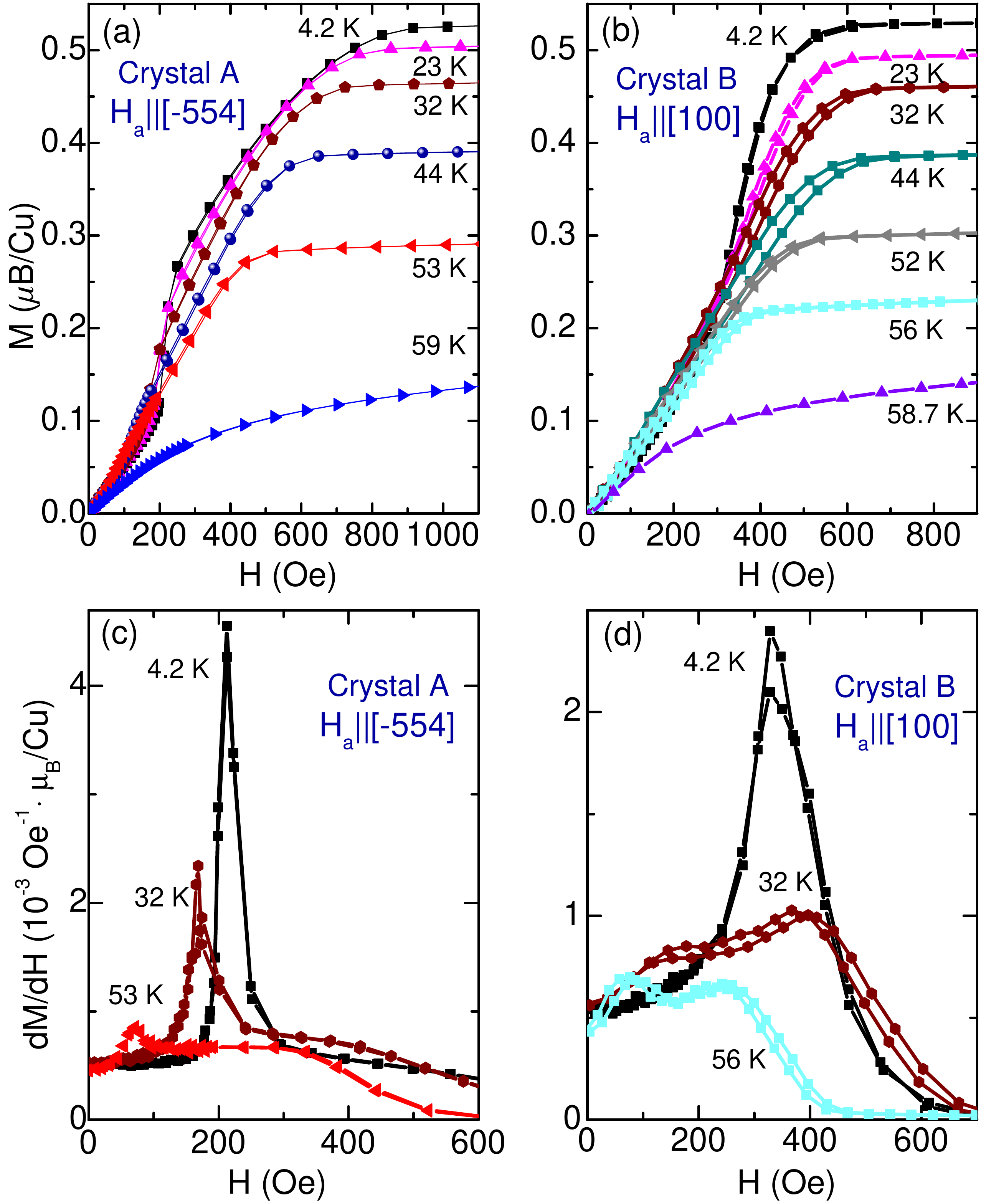}
\caption{(Color online)  M-H curves collected for a number of temperatures and 
with the applied magnetic field H$_a$ along the [-554] (a)  and [100] (b) crystallographic directions.
Panels (c) and (d) show the derivative dM/dH for three representative curves in (a) and (b) respectively.
Here we have corrected for the demagnetizing field H$_d$, i.e. H=H$_{a}-$H$_{d}$ is the internal field.}
\label{MHDerivDemCor}
\end{figure}

Here we would like to map out the H$_a$-T phase diagram of Cu$_2$OSeO$_3$,
so that we contrast it to the one obtained from dielectric measurements below (Sec.~\ref{sec:MEphasediagram}).
To this end, we have taken detailed magnetization measurements as a function of T and H$_a$.
In addition, we have data for two different crystallographic orientations,
by applying the field $\vec{H}_a$ perpendicular to the widest crystal faces of the two samples,
namely to the [-554] direction for Crystal A and to the [100] direction for Crystal B.

A representative set of data for these two crystallographic orientations is shown in Fig.~\ref{MHloop5K}(a), at 5 K.
At low fields the magnetization varies linearly with H$_a$, while a kink is observed at H$_c\simeq 600-800$ Oe,
which is most clearly seen by a peak in the first derivative of the magnetization (see Fig.~\ref{MHloop5K}(b)).
This peak indicates a field-induced transition, which was originally reported in polycrystalline samples,\cite{Bos}
and was also observed in single crystals as well.\cite{Belesi11,Huang}

At higher values of the magnetic field (H$_a=$1.5-2 kOe at 5 K), the magnetization reaches a value of 0.53 $\mu_B$ per Cu for both crystallographic directions.
This moment is consistent with the 3up-1down ferrimagnetic state that was originally proposed by Bos {\it et al.},\cite{Bos} and recently confirmed by high-field NMR measurements.\cite{Belesi10}
In agreement with previous results,\cite{Belesi11} the magnetization reaches this 1/2-plateau earlier when the field is applied along a cube edge,
as compared to a space diagonal direction,  i.e., H$^{[100]}_{\text{plateau}}<$H$^{[-554]}_{\text{plateau}}$.

\begin{figure}[t]
\includegraphics[width=0.44\textwidth]{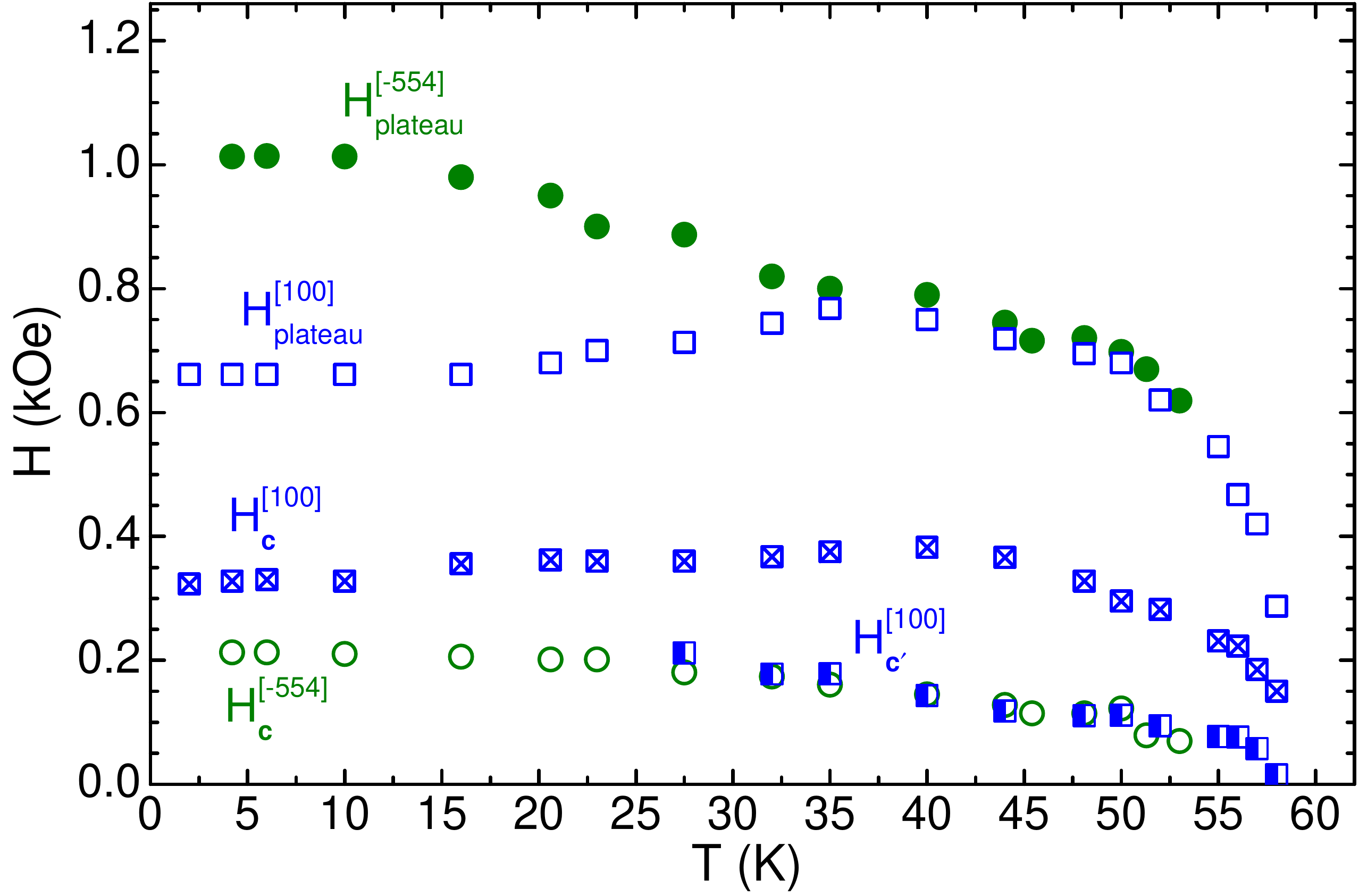}
\caption{(Color online)
The magnetic phase diagram of Cu$_2$OSeO$_3$, as follows from the anomalies in the magnetization measurements shown in Fig.~\ref{MHDerivDemCor}.}
\label{HTdemCorrN2}
\end{figure}

The evolution of the magnetization process with temperature is presented in Figs.~\ref{MHDerivDemCor}(a) and (b),
for the [-554] and [100] crystallographic orientations, respectively.
To highlight some of the features we also show in (c) and (d) the differential susceptibilities dM/dH.
Here, all data are plotted against the internal magnetic field, H=H$_{a}-$H$_{d}$,
where the demagnetizing field H$_{d}$ has been calculated following Ref. [\onlinecite{DemFieldRef}].

To map out the H-T phase diagram of Cu$_2$OSeO$_3$ we follow the T-dependence of the two main characteristic anomalies in the magnetization process,
namely, the position H$_c$ of the field-induced transition and the onset of the 1/2-plateau H$_{\text{plateau}}$.
Collecting H$_c$ and H$_{\text{plateau}}$ for both crystallographic directions, we obtain the H-T phase diagram shown in Fig.~\ref{HTdemCorrN2}.

We first remark that there are two temperature regimes with qualitatively different features, namely above and below 30-35 K.
In the low-T region we find that H$_{c} ^{[-554]} < $H$_{c1} ^{[100]}$, while
H$^{[100]}_{\text{plateau}}<\text{H}^{[-554]}_{\text{plateau}}$, similarly to the 5 K data discussed above.
These findings suggest that the magnetization prefers the body diagonals in the low-field phase, but the [100] axes above the file-induced transition.

In the high-T region, we can first observe that $\text{H}^{[100]}_{\text{plateau}}\simeq\text{H}^{[-554]}_{\text{plateau}}$ and that they both show
an anomalous T-dependence. The second observation is that, in addition to H$_c$ and H$_{\text{plateau}}$,
there appears a third characteristic anomaly, denoted by H$_{c'} ^{[100]}$.
This corresponds to a second peak in dM/dH which is visible for T$>$27 K (Fig.~\ref{MHDerivDemCor}(d)).
It is noteworthy that in the same T-range, the [100] data show a more pronounced hysteresis, which is absent from the [-554] data.
We also note that H$_{c'} ^{[100]}$ follows quite closely the T-dependence of H$_c^{[-554]}$ (see Fig.~\ref{HTdemCorrN2}).
Altogether, the data clearly reveal different anisotropic properties compared to the low-T region, and
suggest either a coexistence region (``phase separation'') between H$_c$ and H$_{c'}$, or a separate phase
whose nature needs to be investigated further.

The magnetic properties in the low-T region are readily interpreted by the
transformations of a ground-state composed of different domains of flat ferrimagnetic helices.
These helices have their propagation directions along the easy axes of the cubic system,
i.e. the magnetic moments are rotating in the plane normal to such an axis.
Since the single-ion magnetic anisotropy is absent for in the present S=1/2 compound,
the leading anisotropy in this cubic system should be of the exchange type.
Depending on the sign of this anisotropic exchange, the easy axes are along [111] or [100].\cite{Bak80,Plumer81}
The ground-state may form a thermodynamical multidomain state, composed of these helices,
owing to magnetoelastic effects, similar to stable domain states in ordinary antiferromagnets.

First-order magnetization processes transform this multidomain state into a single-domain of
a conical helix that propagates in the direction of the applied field and displays a net magnetization.\cite{Plumer81}
The details of this process depend on the orientation of the field with respect to the easy axes and can be hysteretic.
As will be shown below by considering the possible ME coupling terms in the phenomenological theory, for
a cubic magnetoeletric helimagnet like Cu$_2$OSeO$_3$, the
conical helices display a net dielectric polarization, while the effect from the flat helices averages out.
Therefore, the magnetically driven reorientation process may proceed via an intermediate domain
state that is composed of flat helices without a net dielectric polarization and the polarized
field-driven conical helix. In that case, it is the long-range electrostatic effect that stabilizes
a multidomain state of magnetic helices.

The magnetization curves suggest that the transformation process into the single-conical helix
is concluded close to the peak of the $dM/dH$ curves, see Fig.~\ref{MHDerivDemCor}~(b).
For higher fields, the single-domain conical helix evolves toward the collinear ferrimagnetic
state, which is achieved by a continuous process at $H_{\text{plateau}}$.
The T-dependence of $H_{\text{plateau}}$ depicted in Fig.~\ref{HTdemCorrN2}, which
should trace the evolution of the magnetic anisotropy with temperature, is clearly anomalous.
At low temperatures the effective easy axes in the cubic system should be of $[100]$ type at
high fields. As there exists a net dielectric polarization in the conical and collinear phase,
this anisotropy may be affected by the magnetoelectric couplings and may not reflect the purely
magnetic anisotropic couplings which control the behavior of the flat helices at low magnetic fields.
Our survey of the magnetic phase diagram suggests that the magnetic structure is
simple for fields $H>H_c$ as being that of a conical helix that is transformed into a collinear state
by increasing magnetic field at low temperatures, where also a substantially anisotropic magnetic
behavior is observed. However, the magnetic spin-structure and transformation processes at higher temperatures
and close to the magnetic ordering transition will require further experimental and theoretical studies.

\section{Dielectric constant $\varepsilon_r$}\label{sec:er}
The T-dependence of the dielectric constant $\varepsilon_r$ of Cu$_2$OSeO$_3$
has been reported by Bos {\it et al.}\cite{Bos} from capacitance measurements on polycrystalline samples, and
by Miller {\it et al.}\cite{Miller} following an analysis of far-infrared measurements on single crystals.

\begin{figure}[tp]
\includegraphics[width=0.42\textwidth]{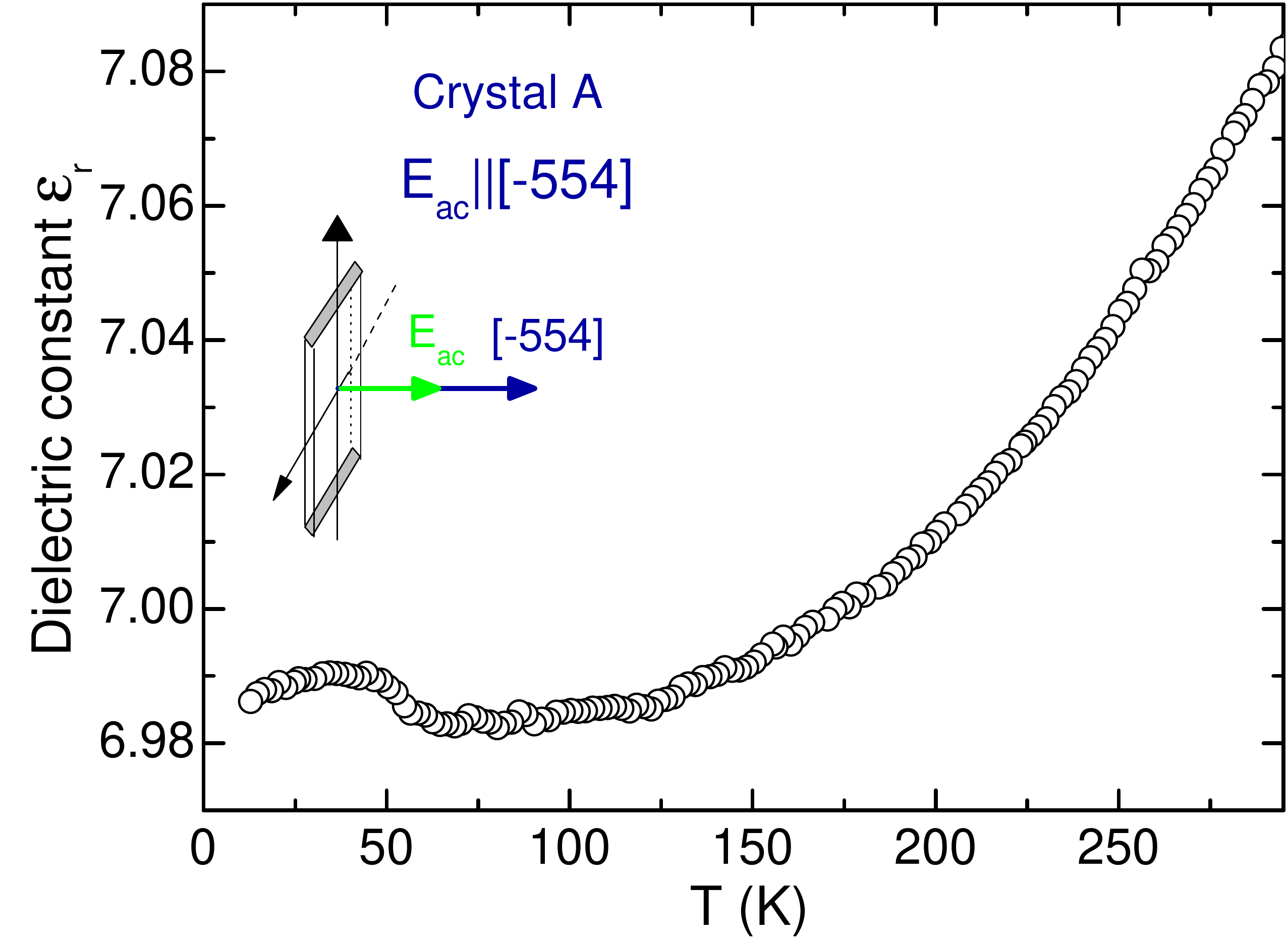}
\caption{ The temperature dependence of the dielectric constant at zero applied magnetic field.}
\label{er111}
\end{figure}

Figure~\ref{er111} shows our own data from capacitance measurements in Crystal A with a parallel plate geometry.
The experiment was performed at zero magnetic field with an ac-electric field along the [-554] direction.
The frequency was 9 kHz and the voltage 1 V. We have repeated the measurement at various frequencies in the range 20 Hz-1 MHz but no significant frequency dependence was observed. The dielectric loss factor ($\tan\delta=\varepsilon'_r/\varepsilon''_r$)  was of the order of $5\times 10^{-4}$ in our measurements. 
The T-dependence of $\varepsilon_r$ was also measured for Crystal B but the results are identical.

Our single-crystal data are in good qualitative agreement with the results reported in Refs.~[\onlinecite{Bos}] and [\onlinecite{Miller}].\footnote{There
are however a few quantitative differences with previously reported data.
First, the absolute values of $\varepsilon _{r}$ are smaller by a factor of two compared to the values reported
in polycrystalline samples,\cite{Bos} and are closer to the ones reported in Ref.~\onlinecite{Miller}.
Second, the drop of $\varepsilon _{r}$ observed for T$<$30 K as well as in the high-T range,
is slower here as compared to the data presented in Ref.~\onlinecite{Bos}.}
The most important feature is the enhancement of the dielectric constant below $\sim$60 K (see Fig.~\ref{er111}),
which marks the onset of an additional contribution which superimposes the high-T curve.
Since this enhancement coincides with the onset of magnetic ordering,
it signals the presence of a magnetoelectric coupling mechanism in Cu$_2$OSeO$_3$.

\section{ME$_H$ effect}\label{sec:MEH}
Here we present direct evidence for the ME$_H$ effect in Cu$_2$OSeO$_3$.
Namely, that an electric charge can be induced at the surfaces of the crystals
by an applied magnetic field (without an applied electric field).
We shall also demonstrate that this phenomenon sets in at the onset of magnetic ordering.

In the following, the measured polarization actually corresponds to $\vec{P}\cdot\vec{e}_s = Q_s/S$, where S is the surface
on which we measure the induced charge $Q_s$, and the unit vector $\vec{e}_s$ is vertical to that surface.

\begin{figure}[tp]
\includegraphics[width=0.4\textwidth]{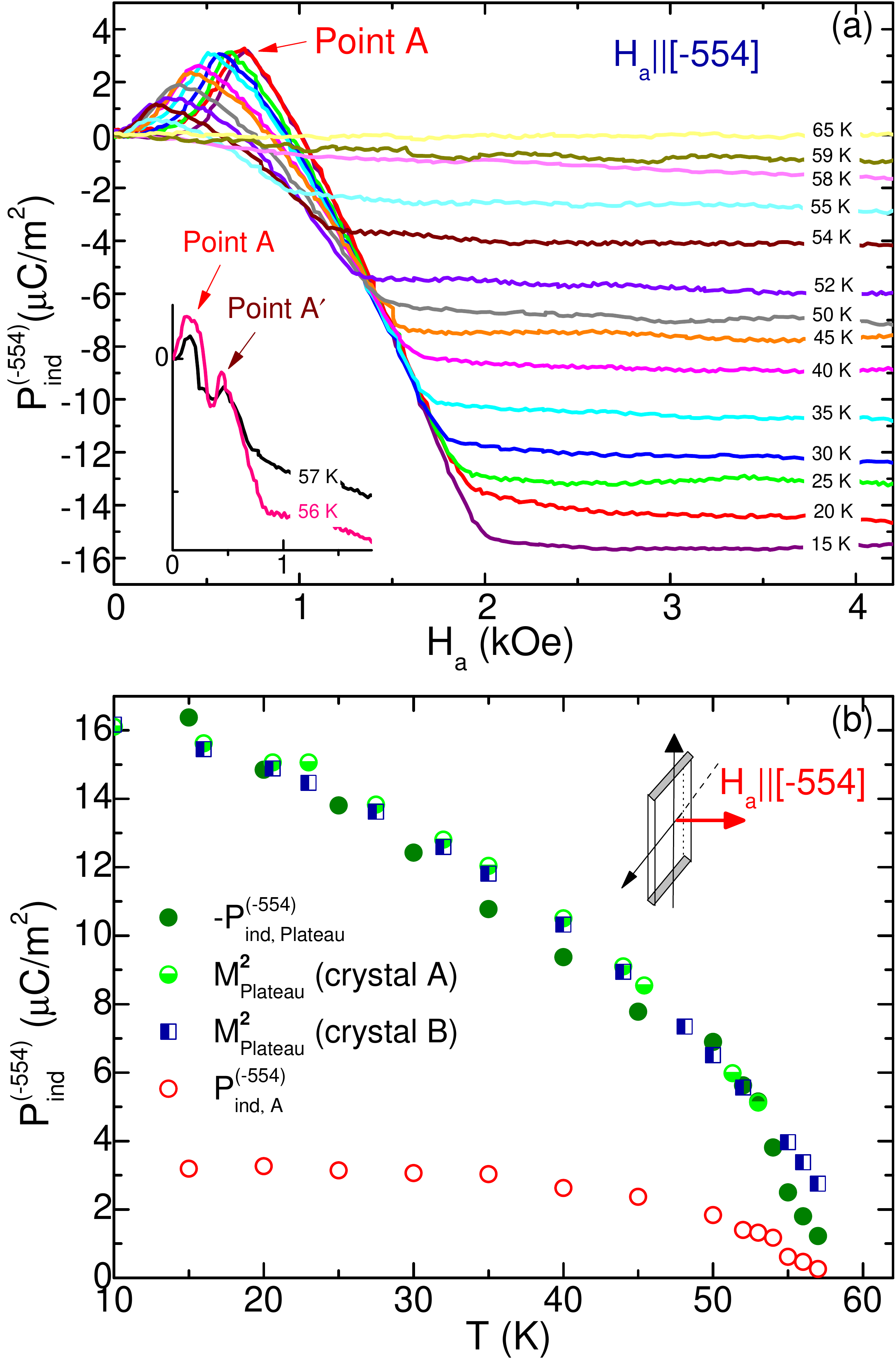}
\caption{(Color online) (a) Magnetic field-induced electric polarization in the [-554] direction at various temperatures.
The magnetic field is also applied along the [-554] orientation.
The inset shows a magnified view of the induced polarization at 56 and 57 K.
(b) The T-dependence of the induced polarization at the position of the field-induced transition
P$^{(-554)}_{\text{ind,A}}$ (open circles), and at the plateau -P$^{(-554)}_{\text{ind,plateau}}$ (filled circles) taken from (a).
Also included the T-dependence of the second power of the magnetization M$^2$ at the position of the plateau for crystal A (half-filled circles)
and crystal B (half-filled squares).}
\label{newP111vsTHpar111}
\end{figure}

\subsection{ME$_H$ effect: T \& H-dependence (``Crystal A'')}\label{sec:TdependenceME}
\subsubsection{$\vec{H}_{a} \parallel [-554],~~\vec{e}_{s} \parallel [-554]$ }
The ME$_H$ effect in Cu$_2$OSeO$_3$ was measured in the range 15-60 K following the quasi-static method described in Section \ref{expdetails}.
Figure~\ref{newP111vsTHpar111}(a) shows the magnetic field induced electric polarization in the [-554] direction for various temperatures.
The magnetic field H$_a$ was applied along the [-554] direction which is perpendicular to the widest crystal face (see inset of Fig.~\ref{newP111vsTHpar111}(b)).

The data at the lowest temperature (15 K) show a net polarization only for H$_a>$ 350 Oe (see also Fig.~\ref{MEdemonstration}).
This finding is consistent with the low-field picture of multi-domain flat helices where the net polarization averages out.
We also observe that the polarization reaches a peak at about 700 Oe, which suggests that the field-induced reorientation of the domains
is concluded at this point.
This is also consistent with the fact that the position of this peak (point ``A'' in the following)
is very close to the field-induced transition in the magnetization measurements (Fig.~\ref{MHloop5K}).

At higher fields, P$^{(-554)}_{\text{ind}}$ drops quickly with H$_a$ and saturates to a negative value
as soon as the system enters the $1/2$-plateau (Fig.~\ref{MHloop5K}). This change of sign between ``point A'' and the onset of the 1/2-plateau
is a very robust feature in our measurements, and, as we show in Sec.~\ref{sec:theoryME},  it is in fact a characteristic fingerprint of the evolution of a
conical helix under a magnetic field.

\begin{figure}[t!]
\includegraphics[width=0.4\textwidth]{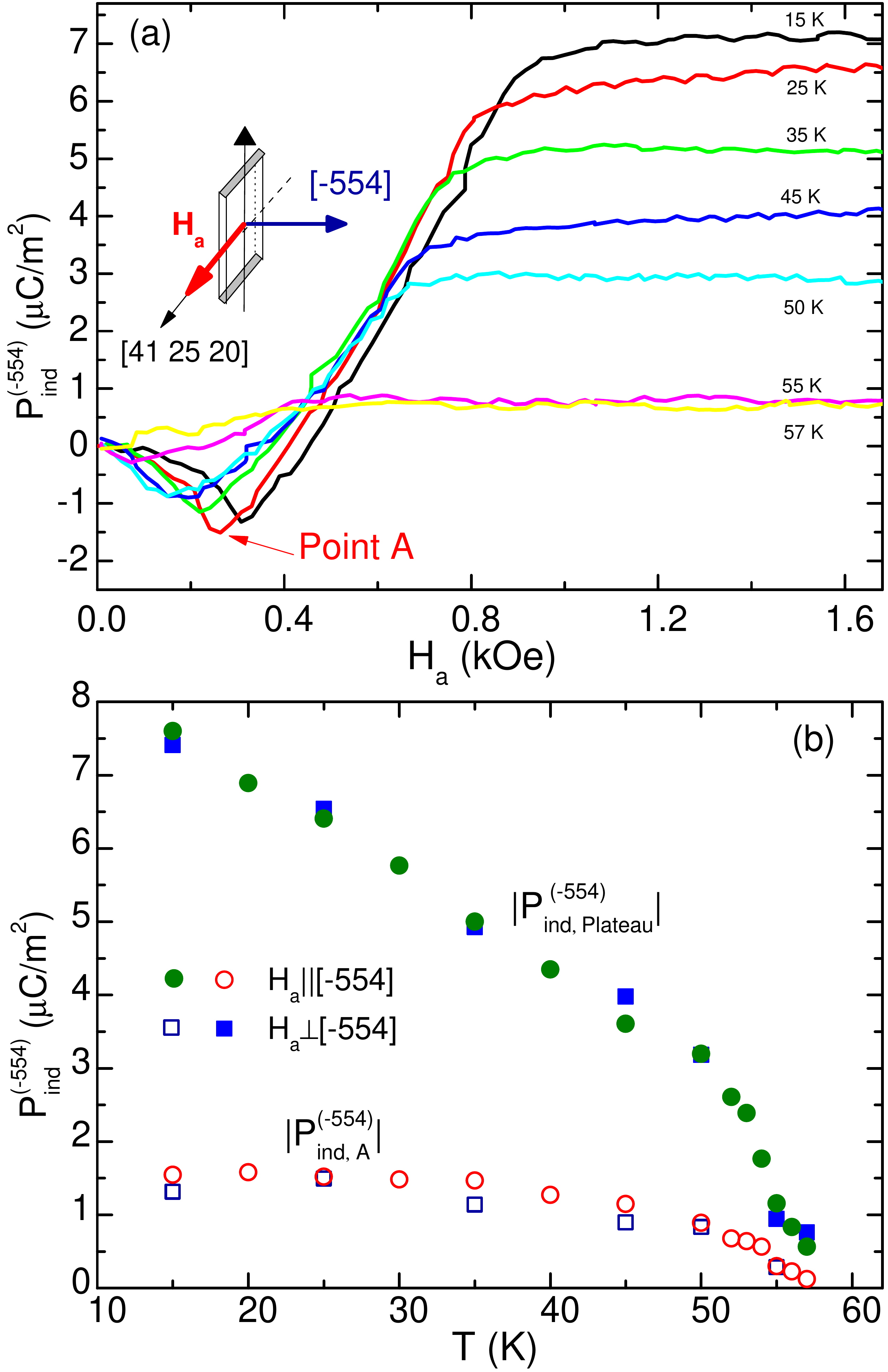}
\caption{(Color online)  (a) Magnetic field-induced polarization in the [-554] direction at various temperatures.
Here the magnetic field is perpendicular to [-554]. (b) The T-dependence of the polarization data shown in (a)
at the field-induced transition P$^{(-554)}_{\text{ind,A}}$ (open squares), and at the plateau P$^{(-554)}_{\text{ind,plateau}}$ (filled squares). We also include the corresponding data (rescaled by a factor of $1/2$) from Fig.~\ref{newP111vsTHpar111}(b) where $\vec{H}_a$ was parallel to [-554] (circles).}
\label{newP111vsTHperp111}
\end{figure}

At higher T's, the overall field dependence remains the same, but the induced polarization is gradually decreasing in magnitude.
A finite polarization can be measured up to 59 K (the anomaly at the field-induced transition can be observed up to 57 K)
which corresponds to the magnetic ordering transition.
This immediately tells us that the magnetoelectric effect is driven by the (primary) magnetic order parameter.

Interestingly, a second anomaly is observed in the P$^{(-554)}_{\text{ind}}$ vs. H curves as we approach T$_c$.
This anomaly (point A$^\prime$ in the inset of Fig.~\ref{newP111vsTHpar111}) could be observed
only between 56-57 K. Despite the fact that the induced charge is rather small in this T-range, this feature was reproducible between different runs.
The existence of this second anomaly may signal the onset of a third phase in a very narrow T-window close to T$_c$, and may well be related to the
presence of the so-called ``A-phase''.\cite{Tokura12}

The T-dependence of the induced polarization at the field-induced transition, P$^{(-554)}_{\text{ind,A}}$,
and at the plateau, P$^{(-554)}_{\text{ind,plateau}}$, are presented in Fig.~\ref{newP111vsTHpar111}(b).
Apart from the sign difference, we can also observe that P$^{(-554)}_{\text{ind,Plateau}}$ shows a much stronger T-dependence than P$^{(-554)}_{\text{ind,A}}$.
Furthermore, we should point out here that the T-dependence of P$^{(-554)}_{\text{ind,plateau}}$ scales quite well with the second power of the magnetization M$^2$ measured
at the position of the plateau (see Fig.~\ref{newP111vsTHpar111}(b)). This happens because the leading order free-energy term responsible for the ME$_H$ effect scales quadratic with the magnetic order parameters (Sec.~\ref{sec:theoryME}).

\subsubsection{$\vec{H}_{a} \perp [-554],~~\vec{e}_{s} \parallel [-554]$ }
We have also studied the T-dependence of P$^{(-554)}_{\text{ind}}$ when $\vec{H}_a\!\!\perp$[-554],
i.e., on the crystal plate (see inset of Fig.~\ref{newP111vsTHperp111}(a)), and the results are presented in Fig.~\ref{newP111vsTHperp111}.
An important difference compared to the data shown in Fig.~\ref{newP111vsTHpar111},
is that here the induced polarization has changed its sign and the magnitude is halved.
This feature will be explained below in Sec.~\ref{sec:theoryME}.
We should remark here that the field-induced transition and the plateau are observed at lower magnetic fields in the measurements of Fig.~\ref{newP111vsTHperp111}(a)
as compared to the data presented in Fig.~\ref{newP111vsTHpar111} (a).
This is simply due to the lower value of the demagnetizing field for the in-plane configuration, as opposed to the out-of-plane orientation shown in Fig.~\ref{newP111vsTHpar111}.

Beside these differences, both P$^{(-554)}_{\text{ind,A}}$ and P$^{(-554)}_{\text{ind,plateau}}$ show
the same T-dependence with that in Fig.~\ref{newP111vsTHpar111}(b).
This is illustrated in Fig.~\ref{newP111vsTHperp111}(b) where we have
included the data from Fig.~\ref{newP111vsTHpar111}(b) (rescaled by a factor of -2).

\subsection{ME$_H$ effect: angular dependence (T=15 K)}\label{AngledependenceME}
We have studied the angular dependence of the ME$_H$ effect for crystals ``A'' and ``B'' at 15 K. The ME$_H$ effect is strongly anisotropic and the direction of the induced polarization can be controlled by rotating the applied magnetic field.

\begin{figure}[tp]
\includegraphics[width=8.6cm]{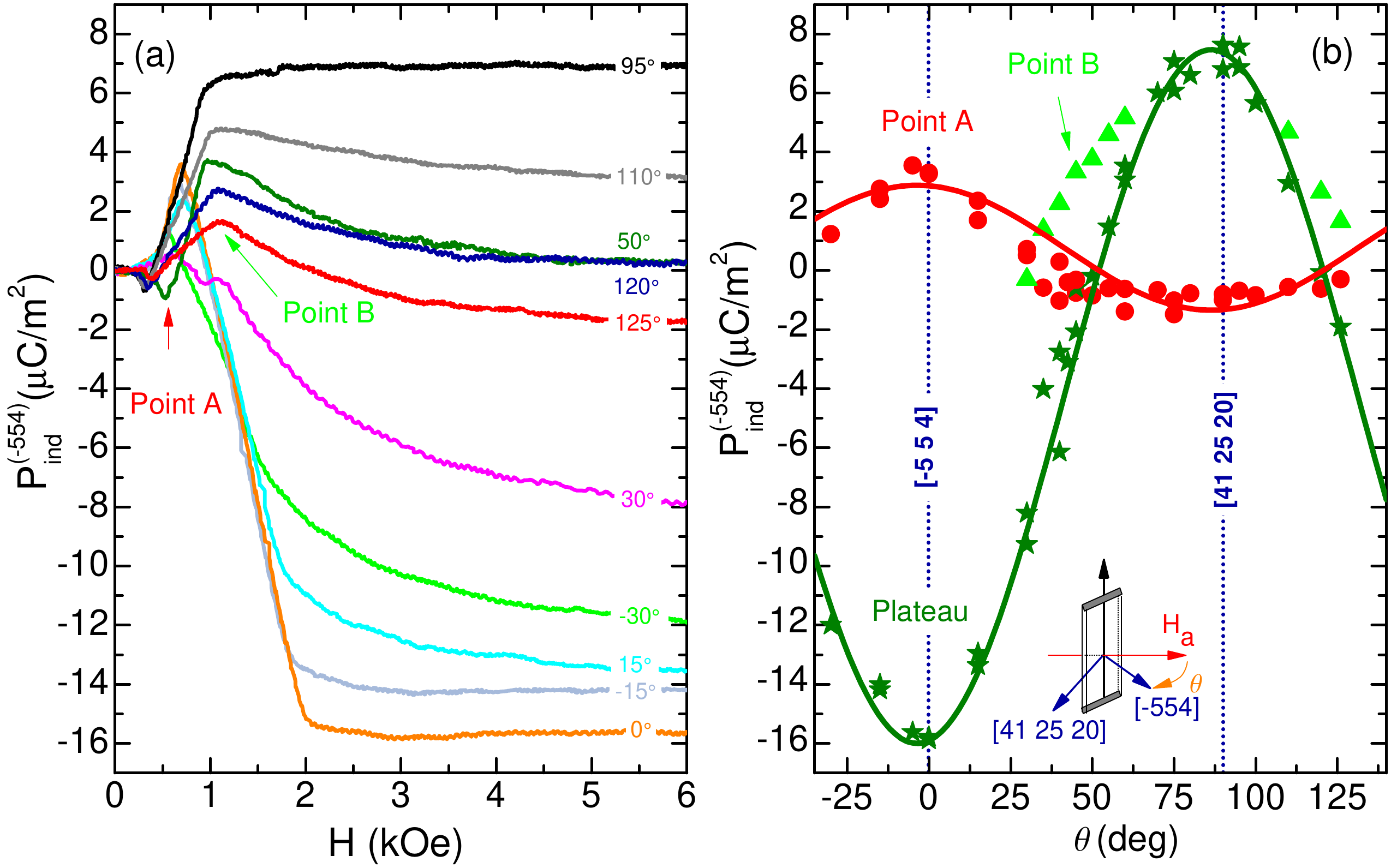}
\caption{(Color online) (a) The magnetic field-induced polarization along [-554], measured at 15 K and for various directions of the magnetic field with  respect to the [-554] axis. At $\theta=0$ the magnetic field is parallel to [-554], while for $\theta=90^\circ$
the magnetic field lies on the (-554) plane. (b) The angular dependence of the induced polarization at the plateau, at ``point A'',
and at ``point B''. The solid lines are fits to Eq.~(\ref{eqn:Akatak}).}\label{PvsTheta111katak}
\end{figure}

\subsubsection{``Crystal A'' }
The angular dependence of the ME$_H$ effect for the (-554) crystal plate is presented in Fig.~\ref{PvsTheta111katak}.
Here we measure the polarization induced along [-554] while we rotate the magnetic field from out-of-plane to in-plane,
as shown in Fig.~\ref{PvsTheta111katak}.

The induced polarization shows an anomaly at the field-induced transition (``point A'') at low fields.
The corresponding amplitude is maximum when the magnetic field is along [-554].
The angular dependence of the induced polarization at ``point A'' and at the 1/2-plateau
can be both captured by our phenomenological theory of Sec.~\ref{sec:theoryME} (solid lines in Fig.~\ref{PvsTheta111katak}(b)).

Apart from the anomaly at ``point A", a second anomaly (marked as ``point B" in Fig.~\ref{PvsTheta111katak}) is observed
between 30$^\circ$ and 60$^\circ$ and between 110$^\circ$ to at least 125$^\circ$. In the region between 60$^\circ$ and 110$^\circ$ this feature cannot be observed
since the field dependence of the induced polarization flattens out and merges with the saturated behavior at the $1/2$-plateau.
The appearance of two anomalies (A and B) must be related to a more complex domain reorientation process in this parameter region,
which requires further investigations.

\begin{figure}[tp]
\includegraphics[width=8.6cm]{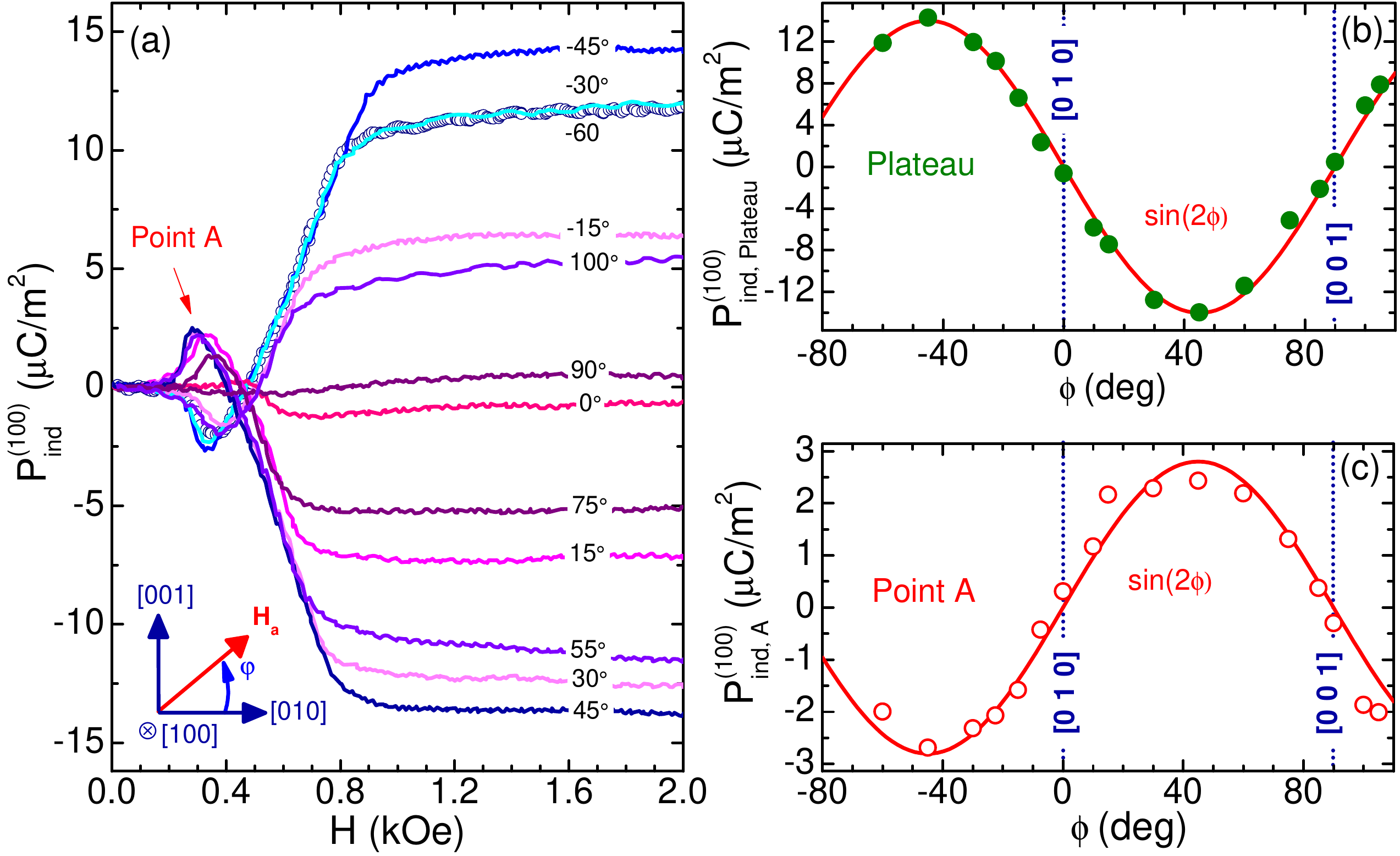}
\caption{(Color online) (a) The magnetic field-induced polarization along [100], measured at 15 K and for $\vec{H}_a$ rotating on the (100) plane
(with $\vec{H}_a \parallel [010]$ at $\phi=0$).
(b)-(c). The angular dependence of the induced polarization at the plateau and at ``point A". The solid lines are fits to Eq.~(\ref{eqn:p100a}).}
\label{MEvsPhi100Horiz}
\end{figure}

\subsubsection{``Crystal B''}
The angular dependence of the ME$_H$ effect for ``Crystal B'' was studied in two different configurations.
In the first configuration (see Fig.~\ref{MEvsPhi100Horiz}(a)) we measure the field-induced polarization along [100]
while we rotate the magnetic field in the (100) crystal plane. For $\phi$=0$^\circ$ the magnetic field is along the [010] axis.
We see that we can switch the direction of the induced electric polarization by rotating the magnetic field in the (100) plane.
In particular, P$^{(100)}_{\text{ind}}$ almost vanishes when the magnetic field is parallel to the cubic edges [010], [001],
and is maximum when it is along the cube face diagonals.

The field dependence of P$^{(100)}_{\text{ind}}$ is similar to the results obtained from the (-554) platelet
(Figs.~\ref{newP111vsTHpar111}, \ref{newP111vsTHperp111}, and \ref{PvsTheta111katak}).
Some finite polarization is observed below the field-induced transition (``point A''), while at higher fields the slope of the P \textit{vs.} H curves
changes sign and finally P$^{(100)}_{\text{ind}}$ saturates as soon as the system enters the 1/2-plateau.
The angular dependence of P$^{(100)}_{\text{ind}}$ at ``point A'' and at the plateau are displayed in Fig.~\ref{MEvsPhi100Horiz}(b) and (c) respectively.
Both curves follow a sinusoidal dependence of the form $\sin(2\phi)$, which can be captured by our phenomenological theory (Sec.~\ref{sec:theoryME}).

\begin{figure}[tp]
\includegraphics[width=8.6cm]{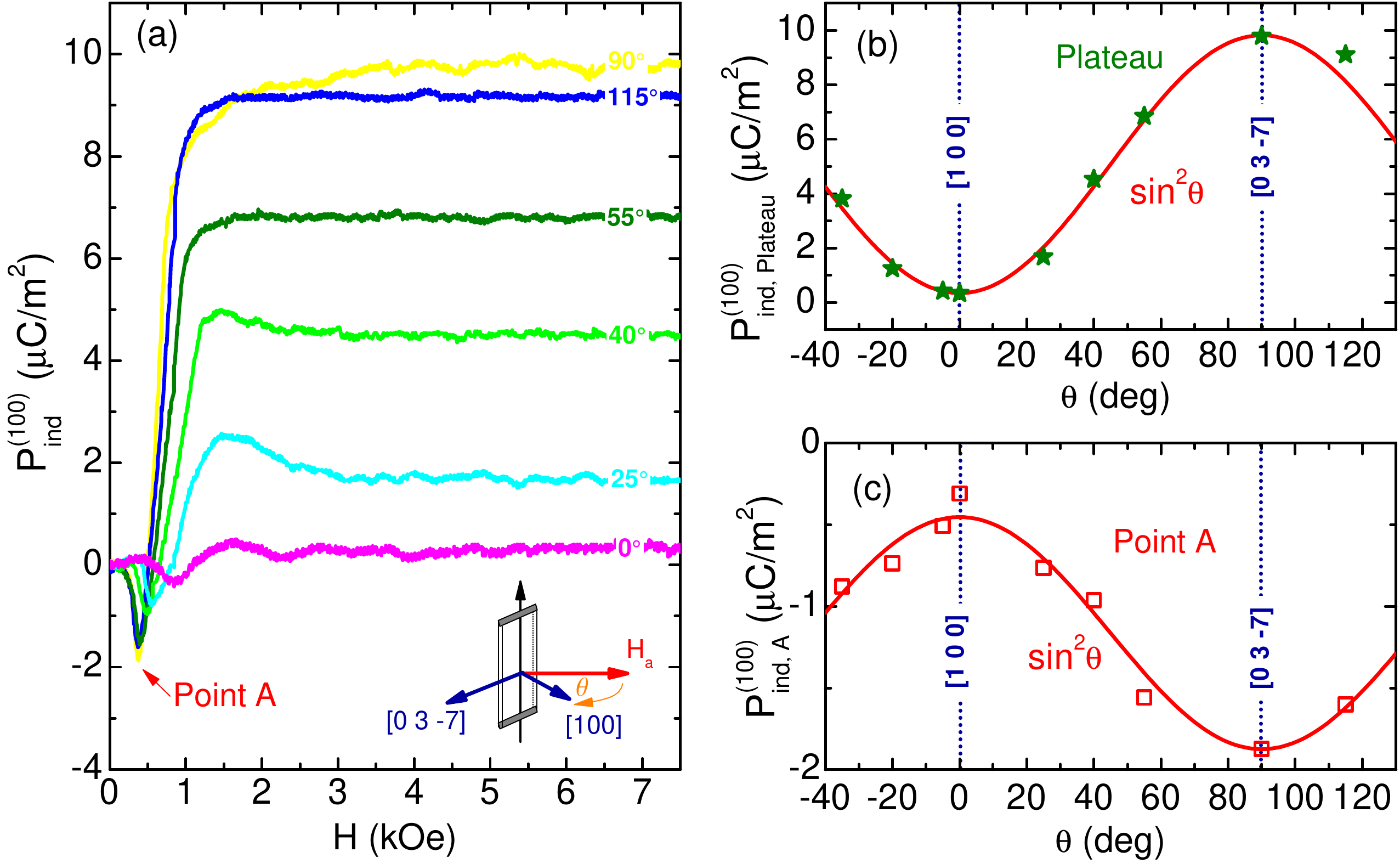}
\caption{(Color online) (a) The magnetic field-induced polarization along [100],  measured at 15 K and for
$\vec{H}_a$ rotating from [100] ($\theta=0$) to [03-7] ($\theta=90^\circ$).
(b) The angular dependence of the induced polarization at the plateau, and (c) at ``point A".
The solid lines are fits to Eq.~\ref{eqn:p100b}.}
\label{PvsTheta100katak}
\end{figure}

In the second configuration, we measure the induced polarization along [100]
while we rotate the magnetic field from out-of-plane to in-plane, as shown in Fig.~\ref{PvsTheta100katak}(a).
The polarization P$^{(100)}_{\text{ind}}$ is almost zero when the field is along [100], and rises with the angle $\theta$
between [100] and the magnetic field. Figure~\ref{PvsTheta100katak} shows the angular dependence of P$^{(100)}_{\text{ind}}$
at the plateau (b) and at ``point A'' (c), and they both scale as $\sin^2\theta$. This dependence
will also be explained in Sec.~\ref{sec:theoryME}.

\section{Magnetocapacitance}\label{sec:MC}
The field dependence of the magnetocapacitance (MC) was first studied in polycrystalline samples of Cu$_2$OSeO$_3$ by Bos {\it et al}.~\cite{Bos}
Here we present our single crystal data for magnetic fields between 0 and 4 T and in the T-range 4.2-60 K.
As we are going to show, the magnetocapacitance
\be\label{MC}
\text{MC}(\text{H}_{a})=\frac{\text{C}(\text{H}_{a})-\text{C}(0)}{\text{C}(0)}
\ee
exhibits a rather strong T-dependence and quite anisotropic properties.
In particular, the magnitude of MC depends on the orientation of the magnetic field with respect to the crystallographic axes
but also on the relative orientation between the magnetic $\vec{H}_a$ and the ac-electric field $\vec{E}_{ac}$.

\begin{figure}[tp]
\includegraphics[width=0.42\textwidth]{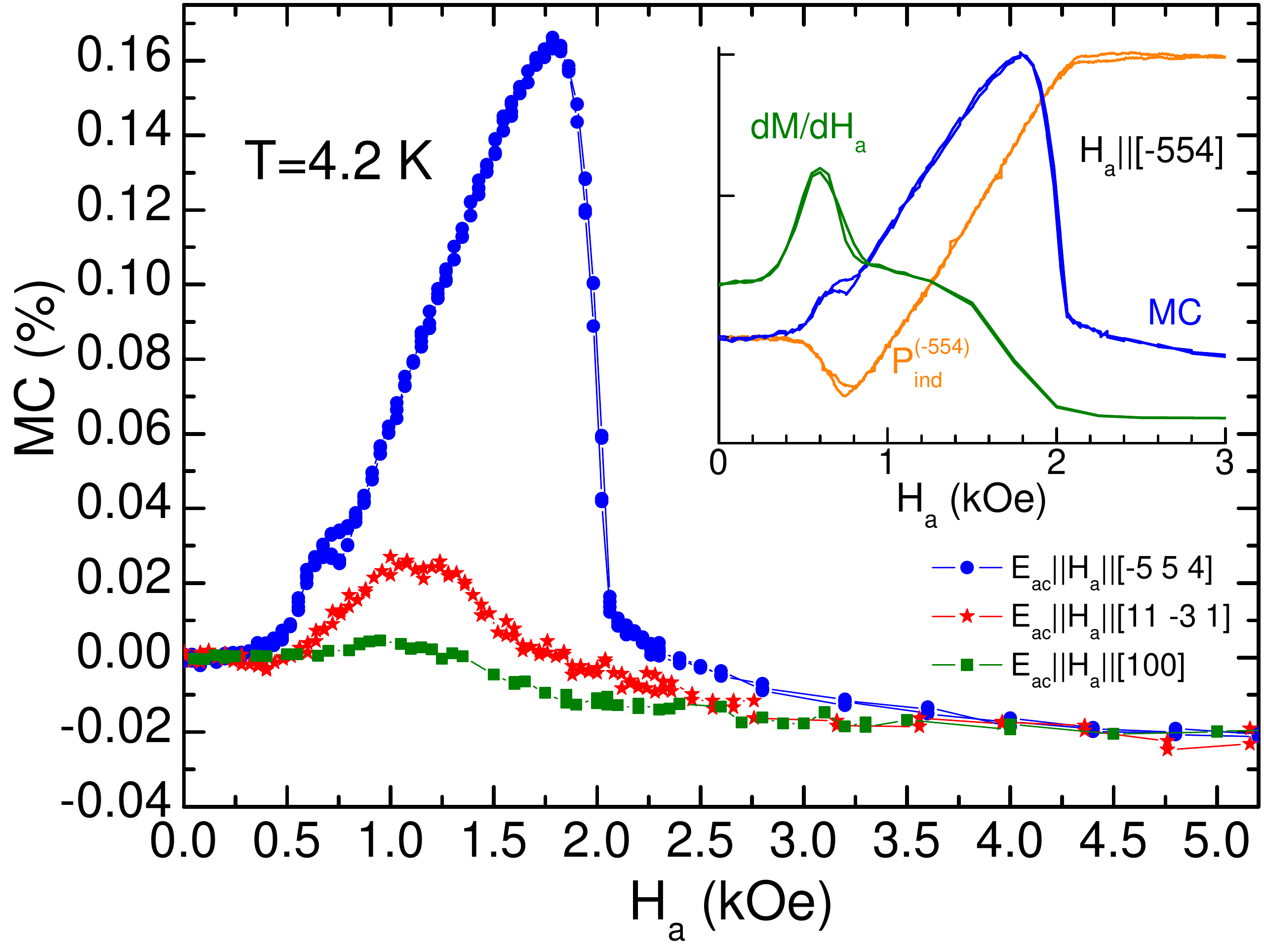}
\caption{(Color online) The magnetocapacitance (MC) of Cu$_2$OSeO$_3$ at 4.2 K
for various orientations of the applied magnetic field H$_{a}$ in respect to the crystallographic axis.
The inset shows the MC data for Crystal A plotted together with the derivative of the magnetization
and the induced polarization P$^{(-554)}_{\text{ind}}$, all measured with the magnetic field applied parallel to [-554].}
\label{MC4p2K}
\end{figure}

\subsection{Representative data with $\vec{H}_{a} \parallel \vec{E}_{ac}$ (T=4.2 K)}
Figure~\ref{MC4p2K} shows some representative MC data at 4.2 K for three different directions of $\vec{H}_a$
with respect to the crystallographic axes, and with $\vec{E}_{ac}\parallel\vec{H}_a$.
Significant MC is observed which is dominated by a peak occurring slightly before the system enters the $1/2$-plateau (see inset of Fig.~\ref{MC4p2K}).
The amplitude of the MC at the position of the peak is maximum when $\vec{H}_a$ is along [-554], and is practically zero when it is directed along [100].
At higher magnetic fields where the system is already in the 1/2-plateau, the MC saturates to small negative values in all directions.
In the inset of Fig.~\ref{MC4p2K}, apart from the MC data for Crystal A we have also included the polarization data P$^{(-554)}_{\text{ind}}$, as well as
dM/dH (from Fig.~\ref{MHloop5K}(b)) for the same crystal, to demonstrate that the anomalies found in the magnetization measurements
track the ones observed in the dielectric measurements, which highlights the strong ME coupling in Cu$_2$OSeO$_3$.

\subsection{T \& H-dependence of MC}
\subsubsection{$\vec{H}_{a} \parallel [-554],~~\vec{E}_{ac} \parallel [-554]$ }
\begin{figure}[tbp]
\includegraphics[width=0.4\textwidth]{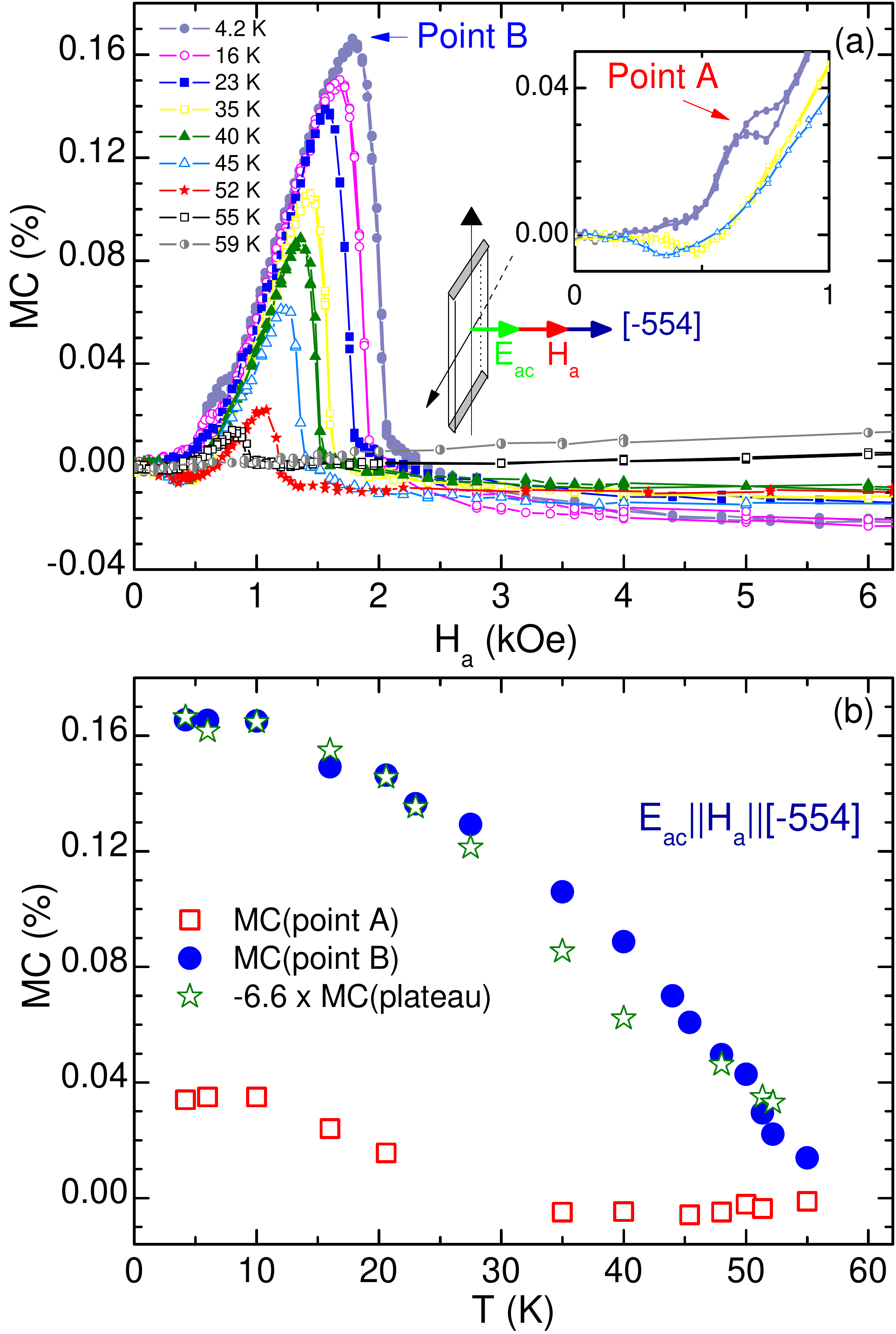}
\caption{(Color online)  (a) The magnetic field dependence of the MC at various T's.
Both magnetic and ac-electric fields were applied along [-554].
The inset shows a magnified view of the MC at 4.2 K at low fields,
where a small shoulder (``point A'') accompanied with some hysteresis is observed.
(b) The magnitude of the MC at three characteristic values of the magnetic field, i.e., at points ``A'' and ``B'',
and at the onset field of the 1/2-plateau (rescaled by a factor of -6.6).}\label{MC111vsTEparH}
\end{figure}

The MC was first studied for the direction exhibiting the maximum effect, namely the [-554] crystal axis.
Figure~\ref{MC111vsTEparH}(a) shows the MC data collected with both $\vec{H}_a$ and $\vec{E}_{ac}$ along [-554].
By increasing T, the MC at the main peak (``point B'') is smoothly decreasing in magnitude
and the position of the peak shifts to lower values of H$_a$.
Apart from the main peak, a small shoulder (``point A'') with some hysteresis is observed at 700 Oe, 
which can also be seen in the inset of Fig.~\ref{MC111vsTEparH}(a).
The position of this shoulder is the same with the field-induced anomalies observed in the ME$_H$
and the magnetization measurements (see inset of Fig.~\ref{MC4p2K}).
The anomaly at ``point A'' is smeared out (and thus it is not discernible) between 20-30 K,
but it can be clearly seen again above 30 K as a small dip and with opposite (negative) MC sign.
We recall here that such a different qualitative behavior above and below the region 20-30 K
was also observed in the [100] magnetization measurements (Fig.~\ref{HTdemCorrN2}),
and thus the two effects must be correlated.
Finally, the MC at the plateau is generally quite small and negative,
but picks up a small positive value above 55 K, which however does not saturate up to 4 T.

Some of the above features can also be seen in Fig.~\ref{MC111vsTEparH}(b),
which shows the MC at ``point A'', at ``point B'', and at the onset of the 1/2-plateau (rescaled by a factor of -6.6).
It is noteworthy that the T-dependence of the MC at ``point B'' is very similar to the one at the position of the plateau.

\begin{figure}[tp]
\includegraphics[width=0.4\textwidth]{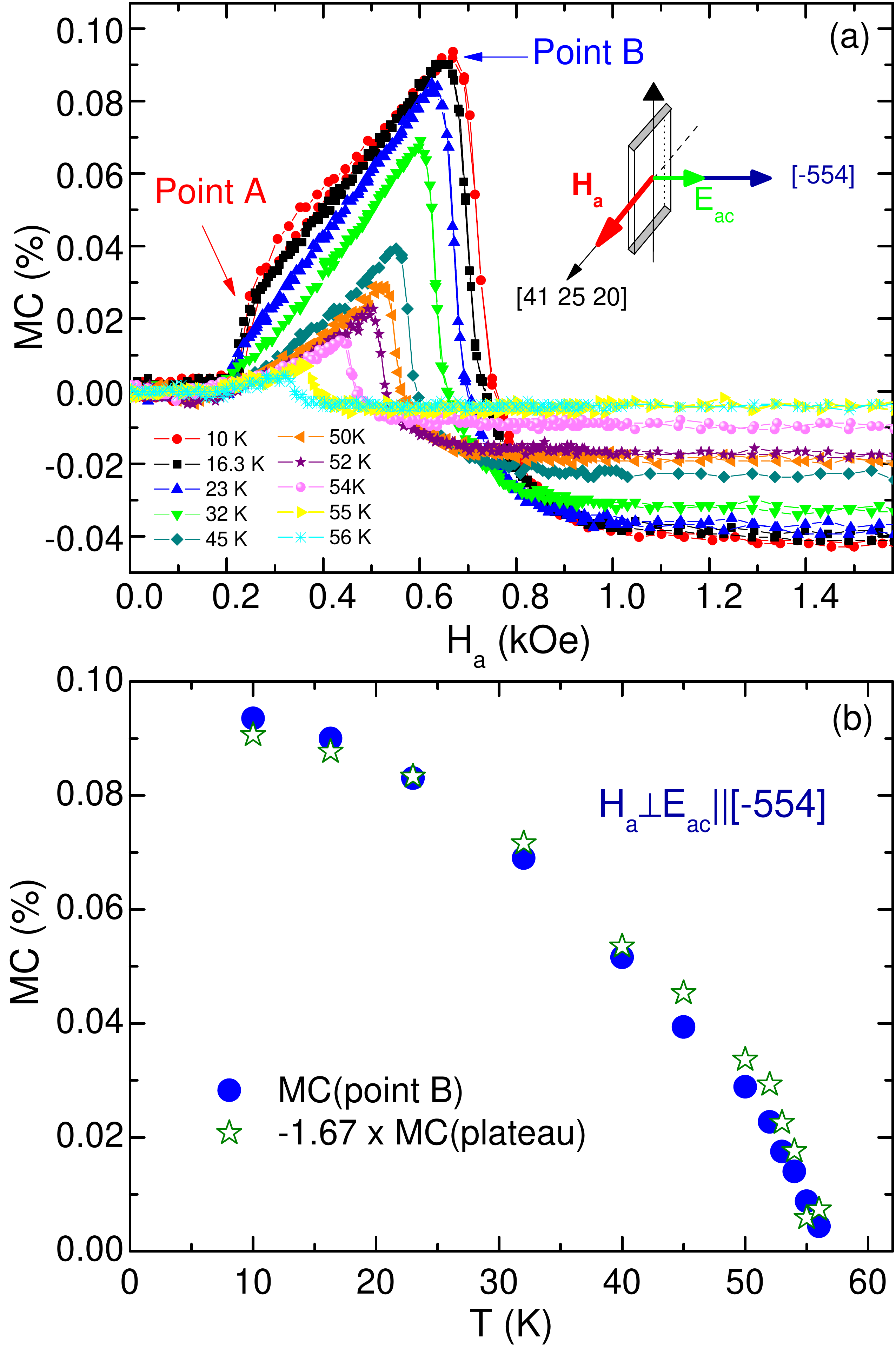}
\caption{(Color online) a) The magnetic field dependence of MC at various temperatures.
The ac-electric field was applied along the [-554] crystallographic orientation,
while the magnetic field is perpendicular to the electric field.
(b) The T-dependence of the magnitude of the MC at saturation and at the point ``B'', taken from (a).}
\label{MC111vsTEperpH}
\end{figure}

\subsubsection{$\vec{H}_{a} \perp [-554],~~\vec{E}_{ac} \parallel [-554]$ }
The MC was also measured with $\vec{H}_a$ in the (-554) plane.
Here the ac-electric field $\vec{E}_{ac}$ is kept in the [-554] direction, as shown in Fig.~\ref{MC111vsTEperpH}.
As in Fig.~\ref{MC111vsTEparH}, the MC is again dominated by a peak (``point B'' in Fig.~\ref{MC111vsTEperpH})
slightly before the onset of the 1/2-plateau, where a negative MC is observed.
At low fields (250 Oe) and for $T\leq$ 23 K, a step-like behavior is observed in the MC data (``point A'').
The field position of this step coincides with the field-induced transition observed in our magnetization measurements (not shown here).
This low-field step-like shape of the MC curves is smeared out for T$>$ 23 K.

The T-dependence of the MC at ``point B'' and at the position of the plateau is presented in Fig.~\ref{MC111vsTEperpH}(b).
The overall dependence is very similar to the data shown in Fig.~\ref{MC111vsTEparH}.
However, the MC at ``point B'' is lower for the in-plane configuration,
while larger (negative) MC values are now observed at the plateau.

\subsection{Angular dependence of MC (T=15 K)}\label{AngledependenceMC}
\begin{figure}[!tp]
\includegraphics[width=8.6cm]{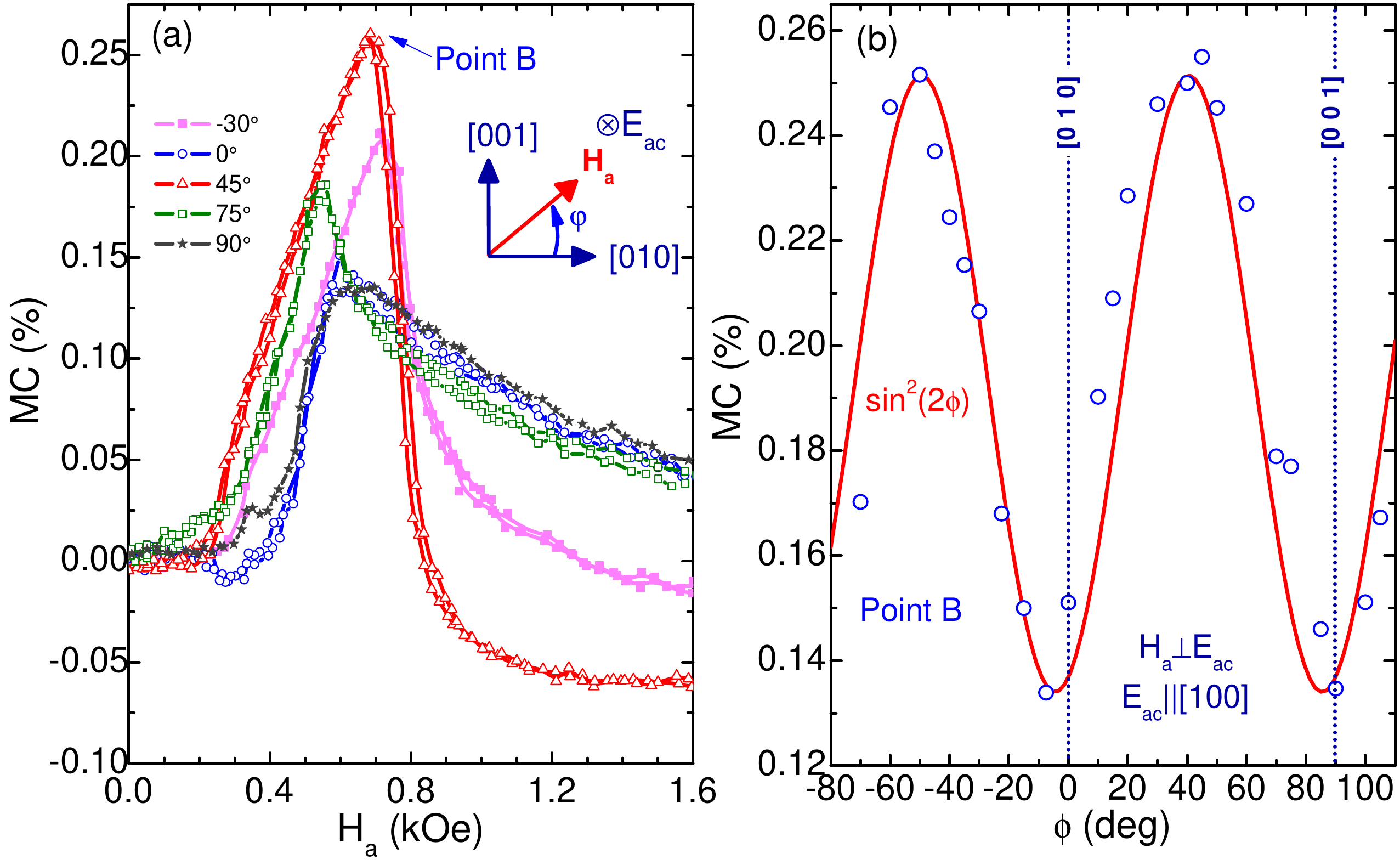}
\caption{(Color online) The angular dependence of the MC, measured at 15 K with the ac-electric field along [100]
and the magnetic field rotating on the (100) plane (at $\phi=0$, $\vec{H}_a\parallel [010]$.
(b) The angular dependence of the magnetocapacitance at ``Point B" is given (b).}
\label{MC100vsPhiHor}
\end{figure}

We have studied the angular dependence of the MC at 15 K for ``Crystal B'' in two different configurations.
In the first configuration shown in Fig.~\ref{MC100vsPhiHor}, $\vec{E}_{ac}$ is along [100] while $\vec{H}_a$ rotates in the (100) plane.
Both the amplitude and the shape of the MC curves change drastically upon changing the orientation of $\vec{H}_a$.
As in the ME$_H$ experiments (Fig.~\ref{MEvsPhi100Horiz}), the maximum effect is observed when $\vec{H}_a$
is parallel to the cube face diagonals. On the other hand, the minimum MC value is observed when $\vec{H}_a$ is along the cube edges.
The MC at the position of the maximum (``point B'') is shown in Fig.~\ref{MC100vsPhiHor}(b) as a function of the angle
$\phi$ between $\vec{H}_a$ and the [010] direction. A sinusoidal angular dependence of the form $\sim\sin^2(2\phi)$ is observed which
suggests a free-energy term of the form $\sim P^2 M^4$ which is quartic in the magnetic order parameter.
As we explain in Sec.~\ref{sec:theoryMC}, such a term arises naturally in the analytical expansion of the free-energy
in powers of the ME coupling mechanism that is related to the ME$_H$ effect.

\begin{figure}[!tb]
\includegraphics[width=8.6cm]{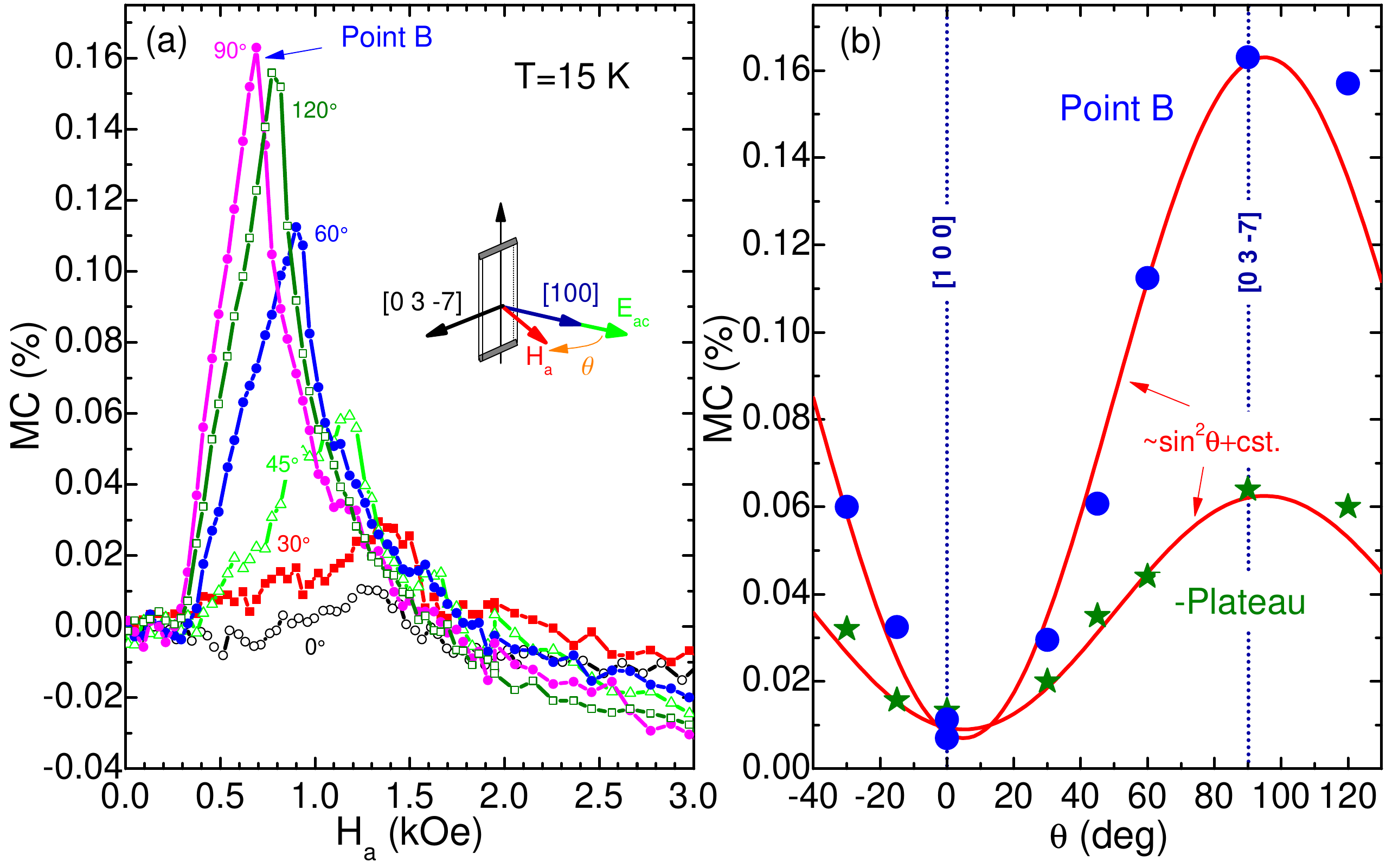}
\caption{(Color online) (a) The angular dependence of the MC measured at 15 K, with
the ac-electric field along [100], and $\vec{H}_a$ rotating from [100] ($\theta=0$) to [03-7] ($\theta=90^\circ$).
(b) The angular dependence of the MC at the plateau and at point ``A'', as taken from (a).}
\label{MCvsAngle100katak}
\end{figure}

In this second configuration, $\vec{E}_{ac}$ is fixed along [100] while $\vec{H}_a$ makes an out-of-plane rotation as shown in Fig.~\ref{MCvsAngle100katak}.
The MC follows a sinusoidal angular dependence of the form $\sim\sin^2\theta$,
both at ``point B" and at the 1/2-plateau. As we discuss in Sec.~\ref{sec:theoryMC}, this dependence suggests the presence of the more conventional
$\sim P^2 M^2$ free-energy term in the free energy.
Finally, it is quite interesting to contrast the large (and positive) MC measured e.g. at ``point B'' when $\vec{H}_a\perp\vec{E}_{ac}$ (first configuration)
to the vanishing MC when $\vec{H}_a\parallel\vec{E}_{ac}$ (second configuration).

\section{Phase diagram from magnetization \& dielectric measurements}\label{sec:MEphasediagram}
\begin{figure}[tp]
\includegraphics[width=0.4\textwidth]{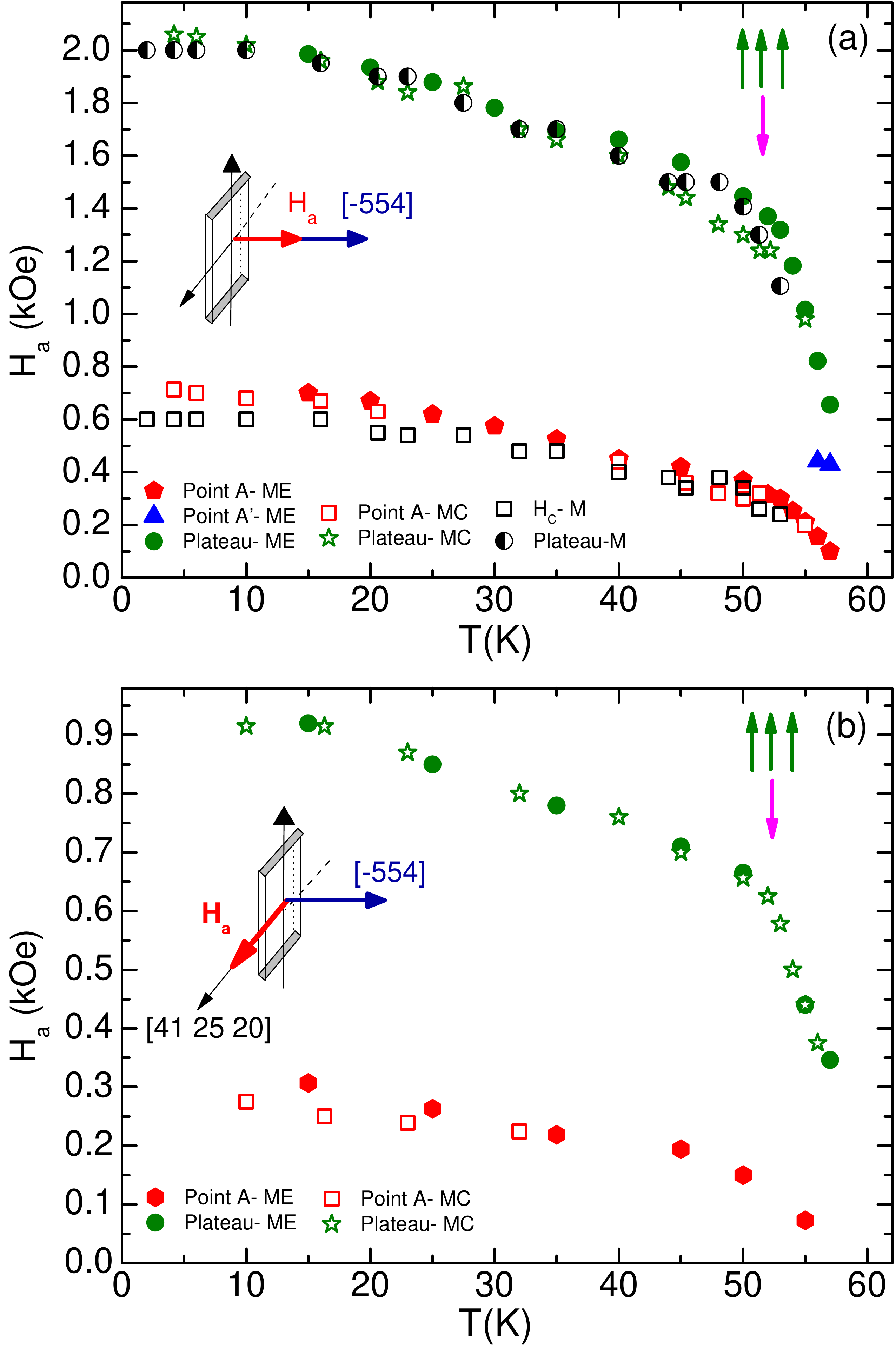}
\caption{(Color online) (a) T-dependence of the field position of the characteristic points ``A", ``A$^{'}$", ``B", and the onset of the 1/2-plateau,
as observed in the field-induced polarization (Fig.~\ref{newP111vsTHpar111}(a)) and MC measurements (Fig.~\ref{MC111vsTEparH} (a)).
(b) Same as in (a), but the results are now from the polarization data of Fig.~\ref{newP111vsTHperp111}(a)
and the MC data of Fig.~\ref{MC111vsTEperpH}(a).}
\label{HcVsTMCME}
\end{figure}

From the position of the characteristic anomalies observed in the ME$_H$ (Figs.~\ref{newP111vsTHpar111} and \ref{newP111vsTHperp111})
and MC (Figs.~\ref{MC111vsTEparH} and \ref{MC111vsTEperpH}) measurements we have deduced the T-H$_a$ phase diagram (Fig.~\ref{HcVsTMCME})
of Cu$_2$OSeO$_3$ for both in and out-of-the (-554) plane orientations of the magnetic field. For the out-of-plane configuration
we have also included the data points from our magnetization measurements for comparison.
The field position at which the low field-induced transition and the plateau are probed from the dielectric measurements fits quite well
with the corresponding positions from the magnetization measurements.
The main differences between Fig.~\ref{HcVsTMCME}(b) and Fig.~\ref{HcVsTMCME}(a)
are due to the different demagnetizing field for the in-plane and the out-of-plane configuration.

Altogether, the dielectric measurements are consistent with the physical picture we described in Sec.~\ref{Magnetization} above, namely that we have at least two
additional phases below the 1/2-plateau, the low-field ($\text{H}\!\!<\!\!\text{H}_c$) and the intermediate-field ($\text{H}_c\!<\!\text{H}\!<\!\text{H}_\text{plateau}$) phase.
The low-field phase is a multi-domain chiral phase whose ME response
defines two sub-regions (not specified in Fig.~\ref{HcVsTMCME}):
The first shows a vanishing ME$_H$ effect which is consistent with the picture of multi-domain averaging of flat helices,
while the second region marks a complex field-induced reorientation process which concludes at the ``point A'' peak anomalies.

The intermediate-field phase is related to the evolution of a single-domain conical helix under a magnetic field, with the propagation vector along the field.
In fact, the phenomenological theory presented in Secs.~\ref{sec:theoryME} and \ref{sec:theoryMC} below, shall provide very strong arguments in favor of this picture.

In addition to the above phases, there are indications in our data for at least one more phase.
For example, the second anomaly  (point A'  in Fig.~\ref{HcVsTMCME}(a))
observed in the small T-window (2K) close to T$_c$ might be related to the presence of the so-called ``A-phase'' that is found in similar compounds.\cite{Wilhelm11,Roessler11,Wilhelm12,Kusaka76,Komatsubara77,Kadowaki82,Ishikawa84,
Gregory92,Lebech95,Thessieu97,Lamago06,Grigoriev06,Grigoriev06a,Muehlbauer09,Neubauer09,Bauer10,
Stishov08,Petrova09,Pappas09,Pappas11,Pfleiderer04,Pedrazzini07,Grigoriev10,Grigoriev11}

\section{Landau-Ginzburg theory for the ME$_H$ effect}\label{sec:theoryME}
\subsection{Conical helix with $\vec{k}\parallel\vec{e}_h$}
Here we develop a phenomenological Landau-Ginzburg approach to account for the ME$_H$ effect.
We shall focus on the intermediate-field and the 1/2-plateau phases.
In this region, the magnetic state can be described by a single-domain conical helix with propagation vector $\vec{k}$ along the field direction $\vec{e}_h$.
We shall keep track of both Cu1 and Cu2 magnetic order parameters by writing
\bea\label{eqn:OPs}
\bs{\eta}_{i}(\vec{r}_i) =\eta_i^\parallel \vec{e}_h + \eta_i^\perp \left( \bs{\psi} e^{i \vec{k}\cdot\vec{r}_i} + h.c. \right)
\eea
with $i=1,2$.
Here $\bs{\psi}=(\vec{e}_1-i\vec{e}_2)/2$, and $\{\vec{e}_1, \vec{e}_2, \vec{e}_h\}$ form a triad
with $\vec{e}_1\times\vec{e}_2 = \vec{e}_h$.
Since the Cu1 and Cu2 moments are antiparallel,
$\eta_1^\parallel=-\eta_1^\parallel\equiv\eta_\parallel$ and similarly $\eta_2^\perp=-\eta_1^\perp\equiv\eta_\perp$.
The long-wavelength transverse twisting between two nearby moments $i$ and $j$
is captured by the relative phase factor $e^{i \vec{k}\cdot\delta\vec{r}_{ij}}$
(where $\delta\vec{r}_{ij}\equiv\vec{r}_i-\vec{r}_j)$, with $\vec{k}\!\cdot\!\delta\vec{r}_{ij}\ll 1$.

\begin{table*}[!t]
\caption{\label{symmetries} Symmetry transformations under the three point group generators of P2$_1$3 and the time reversal $T$. Here $\vec{P}$ is the electric polarization, while $\bs{\eta}$ stands for any of the magnetic order parameters of the system.}
\begin{ruledtabular}\begin{tabular}{c | c |c |c | c}
 & $\{ C_{2z} | (\frac{1}{2},0,\frac{1}{2}) \}$  & $ \{ C_{2y} | (0,\frac{1}{2},\frac{1}{2}) \}$  &  $\{ C_3 | 0 \}$ & ${\it T}$ \\
\hline
$(\eta_x, \eta_y, \eta_z)$ & $(-\eta_x, -\eta_y, +\eta_z)$ & $(-\eta_x, +\eta_y, -\eta_z)$ & $(\eta_z, \eta_x, \eta_y)$ & $(-\eta_x, -\eta_y, -\eta_z)$ \\
\hline
$( P_x, P_y, P_z )$ & $( -P_x, -P_y, +P_z )$ & $( -P_x, +P_y, -P_z )$ & $( P_z, P_x, P_y )$  & $( P_x, P_y, P_z )$ \\
\hline
$( \partial_x, \partial_y, \partial_z )$ & $( -\partial_x, -\partial_y, +\partial_z )$ & $( -\partial_x, +\partial_y, -\partial_z )$ & $( \partial_z, \partial_x, \partial_y )$  & $( \partial_x, \partial_y, \partial_z )$
\end{tabular}\end{ruledtabular}
\end{table*}

We continue by recognizing that the magnetic and electric polarizations of the system
form three-dimensional (3D) representations of the present cubic group $P2_13$, but
the two quantities transform differently under time-reversal.
Table \ref{symmetries} summarizes the transformation properties under the three point-group
generators of $P2_13$ plus the time-reversal invariance.
It is also useful for what follows to define the following composite 3D representations of P2$_1$3
\bea
&& \{ \vec{A},\vec{B} \}_+ = (A^yB^z , A^zB^x, A^xB^y)\\
&& \{ \vec{A},\vec{B} \}_- = \{ \vec{B},\vec{A} \}_+=(A^zB^y , A^xB^z, A^yB^x)
\eea
given two vectors $\vec{A}$ and $\vec{B}$.

Now, to leading order we may look at the invariant local terms that are linear in $\vec{P}$
and quadratic in the magnetic order parameters.
In particular, the theory may contain any pair among $\bs{\eta}_1$ and $\bs{\eta}_2$.
For any such pair ($i$ and $j$, in the following), there are only two invariants that do not involve spatial derivatives (the latter are described in App.~\ref{app:ME}):
\bea
&& \eta_i^y \eta_j^z P^x + \eta_i^z \eta_j^x P^y + \eta_i^x \eta_j^y P^z \equiv
\{ \bs{\eta}_i, \bs{\eta}_j \}_+ \cdot \vec{P}\\
&& \eta_i^z \eta_j^y P^x + \eta_i^x \eta_j^z P^y + \eta_i^y \eta_j^x P^z  \equiv
\{\bs{\eta}_j, \bs{\eta}_i\}_+ \cdot \vec{P}
\eea
where $\{ \bs{\eta}_i, \bs{\eta}_j \}_{\pm}$ are two composite 3D polar representations of $P2_13$, as defined above.
So to lowest order, the free energy reads
\bea\label{eqn:fe}
\Phi(\{\bs{\eta}\},\vec{P}) &=&\Phi_0(\{\bs{\eta}\}) + \frac{\vec{P}^2}{2\chi_e} - \vec{P} \cdot \vec{E} \nonumber\\
&-& \vec{P} \cdot \Big( \xi_1 \{ \bs{\eta}_i, \bs{\eta}_j \}_+ + \xi_2 \{ \bs{\eta}_i, \bs{\eta}_j \}_- \Big) ~.
\eea
Here $\Phi_{0}$ stands for the magnetic portion of the energy in the absence of the ME coupling (i.e. it contains the strong exchange
energy plus the chiral DM term and other types of anisotropies),
$\chi_e$ is the electric susceptibility (scalar for cubic groups) in the absence of ME coupling, and
$\xi_{1,2}$ are the ME coupling constants.  The physical origin of the last term of Eq.~(\ref{eqn:fe}) becomes more transparent by rewriting it as
\be
\frac{1}{2} \vec{P}\cdot \Big[ \xi_+ \left( \{ \bs{\eta}_i,\bs{\eta}_j\}_+ + \{ \bs{\eta}_i,\bs{\eta}_j\}_- \right)
+ \xi_- \bs{\eta}_i \times \bs{\eta}_j \Big] ~,
\ee
where $\xi_\pm \equiv \xi_1 \pm \xi_2$. The first term inside the brackets involves the symmetric exchange anisotropy,
while the second involves the antisymmetric DM interaction.
These two anisotropy types show drastically different contributions to the ME$_H$ effect. Indeed,
minimizing $\Phi$ with respect to $\vec{P}$ gives  (for $\vec{E}=0$):
\bea\label{eqn:p1}
\vec{P}/\chi_e =  
\frac{\xi_+}{2} \left( \{\bs{\eta}_i,\bs{\eta}_j \}_+ + \{\bs{\eta}_i,\bs{\eta}_j \}_- \right) + \frac{\xi_-}{2} \bs{\eta}_i\times\bs{\eta}_j ~.
\eea
It is now evident that the ME$_H$ effect will be dominated by the symmetric exchange component,
since nearby spins are almost collinear in the long-wavelength conical state.
Thus, despite the fact that the DM coupling is the primary source of helimagnetism, its influence on
the ME$_H$ effect is heavily diminished by its long-wavelength nature.

To make these points more explicit and obtain the final predictions for the angular and field dependence of the ME$_H$ effect,
we proceed by replacing Eq.~(\ref{eqn:OPs}) in Eq.~(\ref{eqn:p1})
One immediate prediction is that the helimagnetism induces both a uniform as well as
a spatially oscillating polarization component which is also of mesoscopic nature.
In the following we shall restrict ourselves to the uniform component, $\vec{P}^{\text{uni}}$,
which is the one we measure in our experiments.
Using the relation $4 \{ \bs{\psi}, \bs{\psi}^\ast \}_\pm =  - \{ \vec{e}_h, \vec{e}_h\}_+ \pm i \vec{e}_h$,
and expanding $e^{i\vec{k}\cdot\delta\vec{r}_{ij}}\simeq 1+ \vec{k}\cdot\delta\vec{r}_{ij}$, we find:
\bea\label{eqn:p2}
\frac{\vec{P}^{\text{uni}}}{\sigma_{ij}\chi_e} &=& \xi_+ \Big( \eta_\parallel^2-\frac{1}{2}\eta_\perp^2\Big)~\{ \vec{e}_h, \vec{e}_h \}_+ \nonumber\\
&-&\frac{\xi_-}{2} \eta_\perp^2 ~(\vec{k}\cdot\delta\vec{r}_{ij})~ \vec{e}_h~
\eea
where $\sigma_{ij}$ is the sign of $\eta_i^\parallel \eta_j^\parallel$.
The second term is the contribution from the DM coupling and can be disregarded since $\vec{k}\cdot\delta\vec{r}_{ij}\ll 1$, as discussed above.
Similar contributions that scale linearly with $k$ arise also from the inhomogeneous Lifshitz-type invariants in the free energy (App.~\ref{app:ME}).
Altogether we may write
\bea\label{eqn:p3}
\frac{\vec{P}^{\text{uni}}}{\sigma_{ij} \chi_e} &\simeq&  \xi_+ \Big( \eta_\parallel^2 - \frac{1}{2}\eta_\perp^2 \Big)~\{ \vec{e}_h, \vec{e}_h \}_+
\eea
which is our central expression.
As we are going to show below, the angular dependence found in the experiments is fully captured by the
projection of the vector $\{\vec{e}_h,\vec{e}_h\}_+$ along $\vec{e}_s$.
Similarly, the observed field dependence can be fully captured (in the field range where the magnetic state
is described by Eq.~(\ref{eqn:OPs})) by the quantity $\eta_\parallel^2 - \frac{1}{2}\eta_\perp^2$.
This quantity goes from $-1/2$ for a completely flat helix ($\eta_\parallel = 0$) to +1 at the 1/2-plateau (where $\eta_\perp=0$),
and thus provides a physically transparent explanation for the sign difference between the measured polarization at ``point A''
and that at the 1/2-plateau. Of course, the ``point A'' does not correspond to a completely flat helix (this would be the case at zero-field)
and this is why the relative magnitude $|P_A|/|P_{\text{plateau}}|$ is actually smaller than 1/2.

\subsection{Flat helix with $\vec{k}\nparallel\vec{e}_h$}
Here we comment what happens for a flat helix whose propagation vector is not along $\vec{e}_h$,
but along some general direction $\vec{e}_k$. This analysis explains why the total polarization averages
out in the low-field region of the multi-domain helical phase.

Given that $\vec{e}_1\times\vec{e}_2$ is now equal to $\vec{e}_k$, one may easily show that the above formulas remain valid
if we replace $\vec{e}_h$ by $\vec{e}_k$ everywhere. In particular, the net polarization is given by
\be
\vec{P}^{\text{uni}} \propto  \sum_{\vec{k}\text{-domains}} \{\vec{e}_k,\vec{e}_k\}_+~.
\ee
Now, the magnetization process close to the 1/2-plateau (Fig.~\ref{MHloop5K}(a))
suggests the $\langle100\rangle$ as the easy axes, but the low-field data cannot confirm this
(in fact they seem to suggest the $\langle111\rangle$ as the easy axes
but this is most likely related to the complex low-field domain reorientation process, as discussed above).
In any case, one can show that the net polarization vanishes for both $\langle100\rangle$ $\langle111\rangle$ cases.
In the first case, $\vec{e}_k\!\parallel\!\langle100\rangle$, all three domains give a vanishing contribution, while
in the second, $\vec{e}_k\!\parallel\!\langle111\rangle$, each of the four domains gives a finite contribution
(for a flat helix $\eta_\parallel^2-\eta_\perp^2/2 = -1/2$) but the net polarization from all domains averages out.

\subsection{ME$_H$ effect: Comparison with exp. data}
\subsubsection{Angular dependence}
We now proceed to a direct comparison to the experimental ME$_H$ data, beggining with the angular dependence.
In the following we shall use the scaled quantity
\be
\tilde{P} = \frac{P}{\sigma_{ij} \chi_e \xi_+ (\eta_\parallel^2-\frac{1}{2}\eta_\perp^2) }~.
\ee
We begin with crystal A for which $\vec{e}_s=\frac{1}{\sqrt{66}} (-5,5,4)$.
For the data shown in Fig.~\ref{PvsTheta111katak}(b), $\vec{e}_h=\cos\theta~\vec{e}_s + \sin\theta~\vec{e}_A'$, where
$\vec{e}_A'=\frac{1}{\sqrt{2706}}(-41, -25, -20)$. Applying Eq.~(\ref{eqn:p2}) we find
\be\label{eqn:Akatak}
\tilde{P}^{(-554)} = -\frac{5}{451}\sqrt{\frac{2}{33}} \left( 55 +150 \cos(2\theta)  - 3\sqrt{41} \sin(2\theta) \right)
\ee
The solid line shown in Fig.~\ref{PvsTheta111katak}(b) is a fit using this expression.
Parenthetically, we should remark that the term proportional to $\sin(2\theta)$ in Eq.~(\ref{eqn:Akatak})
arises from the small  deviation (6$^\circ$) of $\vec{e}_s$ from [-111].
This is why the proportionality constant in front of this term is much smaller than the first two terms of Eq.~(\ref{eqn:Akatak}).
Would $\vec{e}_s$ be exactly along [-111] we would get
\be
\tilde{P}^{(-111)} =- \frac{1+3 \cos(2\theta)}{4\sqrt{3}}~.
\ee
These expressions explain in particular why the polarization data with
$\vec{H}_a \parallel \vec{e}_s$ ($\theta=0$) are larger by a factor of -2 from the data corresponding to
$\vec{H}_a \perp \vec{e}_s$ ($\theta=90^\circ$), see Fig.~\ref{newP111vsTHperp111}.

Let us now turn to Crystal B for which $\vec{e}_s=[100]$.
For the data shown in Fig.~\ref{MEvsPhi100Horiz}, the magnetic field rotates in the plane
defined by $\vec{e}_h=\cos\phi ~[010] + \sin\phi ~ [001]$, and Eq.~(\ref{eqn:p2}) gives
\be\label{eqn:p100a}
\tilde{P}^{(100)} = \frac{1}{2} \sin(2\phi)
\ee
which is also in perfect agreement with the data.

Similarly, for the data shown in Fig.~\ref{PvsTheta100katak}, the magnetic field rotates in the plane
defined by $\vec{e}_h=\cos\theta ~[100] + \sin\theta ~ \vec{e}_B'$, where $\vec{e}_B'=\frac{1}{\sqrt{58}} (0,3,-7)$,
and Eq.~(\ref{eqn:p2}) gives
\be\label{eqn:p100b}
\tilde{P}^{(100)} = -\frac{21}{58} \sin^2\theta~,
\ee
which is again in very good agreement with the data, apart from a very small shift
which may originate either from a tiny misalignment of $\vec{e}_s$ away from [100],
or from additional minor contributions to the ME free energy.

\subsubsection{Field dependence}
Let us now turn to the field-dependence of the measured polarization and compare it with Eq.~(\ref{eqn:p2}).
To this end we shall make use of some well known and rather simple theoretical expressions
for the evolution of $\eta^\parallel$ and $\eta^\perp$ as a function of applied field, in the conical phase.
Following e.g. Ref.~\onlinecite{Kirkpatrick}, we write
\be\label{eqn:kirkpatrick}
\eta_\parallel = g \frac{H}{H_{\text{plateau}}}, ~~\eta_\perp = g \sqrt{ 1-H^2/H_{\text{plateau}}^2}~,
\ee
where $g$ is a T-dependent prefactor which vanishes at $T_c$, and $H_{\text{plateau}}$ is the onset field of the 1/2-plateau.

\begin{figure}[!tp]
\includegraphics[width=0.46\textwidth]{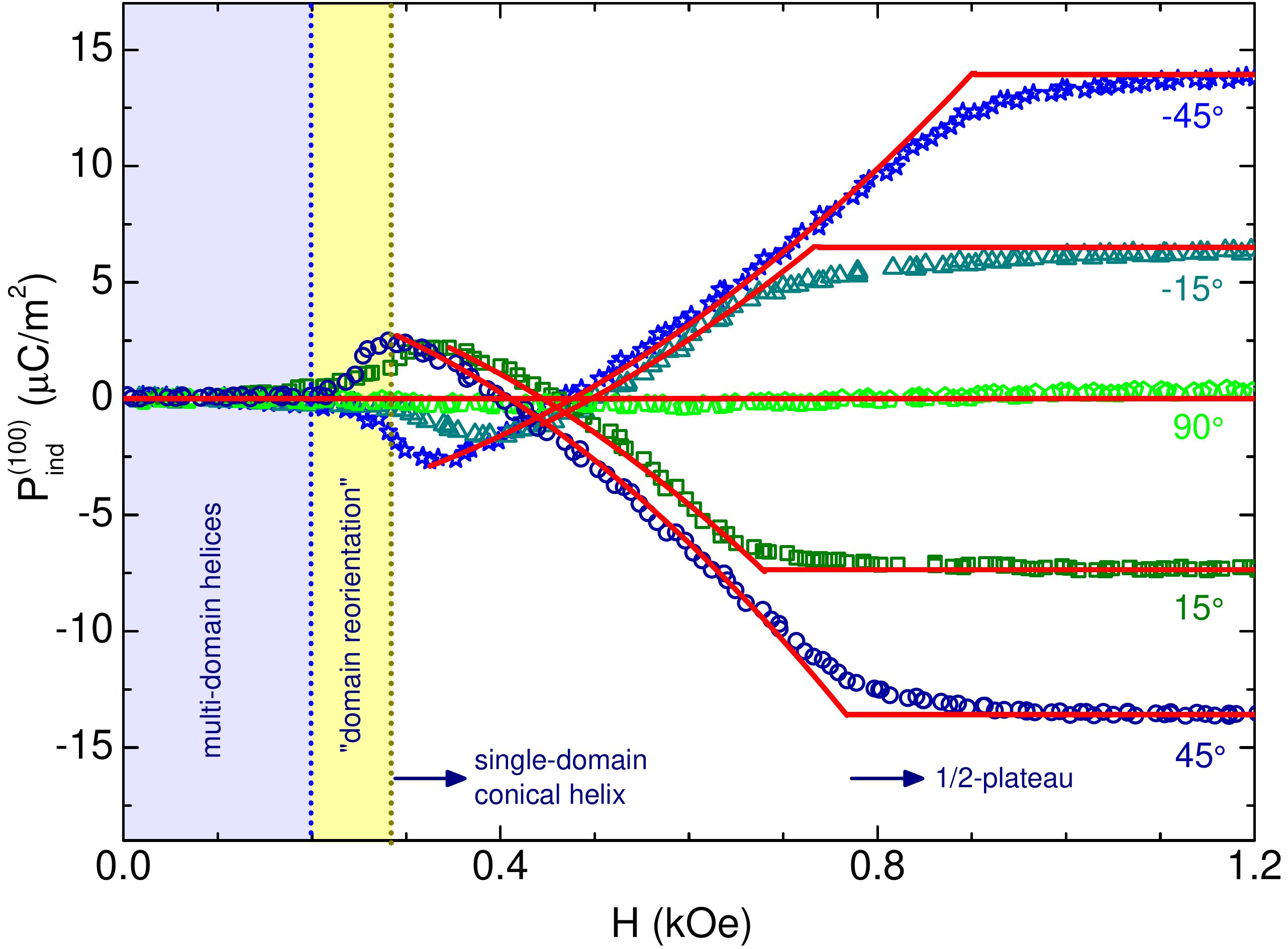}
\caption{(Color online) Comparison between theory (solid lines, see text) and experimental data (symbols)
for the field dependence of the electric polarization along [100] for some representative angles $\theta$
(data reproduced from Fig.~\ref{MEvsPhi100Horiz}(a)). The shaded area shows the low-field phase (which further consists of two sub-regions).
The first arrow designates the onset of the single-domain conical phase (peak at ``point A'') for $\theta=45^\circ$.
This phase evolves continuously toward the 3up-1down ferrimagnetic phase (1/2-plateau). }
\label{fig:P100ExpTheory}
\end{figure}

For concreteness we consider the data shown in Fig.~\ref{MEvsPhi100Horiz}.
According to Eq.~(\ref{eqn:p100a}),
\be
\frac{P^{(100)}}{\sigma_{ij} \chi_e \xi_+ } = \frac{g^2}{4} \Big( 3 \frac{H^2}{H_{\text{plateau}}^2} -1 \Big) ~\sin(2\phi)~,
\ee
where we note that the polarization scales quadratically with $H$.
From our data we have access to both parameters $g$ and $H_{\text{plateau}}$ (the latter is influenced by the demagnetizing field),
and so we can make a direct comparison between theory and experiment.
This comparison is shown in Fig.~\ref{fig:P100ExpTheory} for a few representative angles $\theta$, and is remarkably successful.
The deviation around the 1/2-plateau can be ascribed to thermal fluctuations (the experimental data are taken at 15 K
while the above expression is valid only at T=0).

Altogether, this gives further confidence that the ME coupling mechanism of Eq.~(\ref{eqn:fe}) above captures
the experimental findings both qualitatively and quantitatively.

\section{Landau-Ginzburg theory for the magnetocapacitance}\label{sec:theoryMC}
\subsection{Theory}
To capture the physics of the MC, we must include higher order terms in the ME portion of the free energy, and in particular we must look at
quadratic terms in the polarization. To satisfy time-reversal invariance this term must scale as $(\eta)^p P P$, were $p$ is an even integer.
Now the strong $\sim \sin^2(2\phi)$ angular dependence of Fig.~\ref{MC100vsPhiHor}(b) suggests that we should take $p=4$,
while $p=2$ may also be present as suggested by the angular dependence of Fig.~\ref{MCvsAngle100katak}(b).

The simplest possible term with $p=4$, that is also physically motivated,
is the square of the symmetric combination of the ME$_H$ invariant of Eq.~(\ref{eqn:fe}), namely
\be\label{eqn:mc}
\mc{U}_0\!=\!\frac{1}{4}\left[\left( \{\bs{\eta}_i,\bs{\eta}_j\}_+ +\{\bs{\eta}_i,\bs{\eta}_j\}_- \right)\!\cdot\!\vec{P}\right]^2
\!=\!\left(\{\bs{\eta}_1,\bs{\eta}_1\}_+\!\cdot\!\vec{P}\right)^2,
\ee
where in the second equality we disregarded terms related to the small phase difference $\vec{k}\cdot\delta\vec{r}_{ij}$, as discussed above.
Physically, $\mc{U}_0$ stands for the second-order term in the analytical expansion of the free energy
in the ME coupling mechanism discussed in Sec.~\ref{sec:theoryME}. Including this term gives
\be\label{eqn:feMC1}
\delta\Phi = \Phi'-\Phi = -\zeta_0~\mc{U}_0,
\ee
where $\zeta_0$ is the coupling constant, and $\Phi$ is given in Eq.~(\ref{eqn:fe}) above.

The polarization is again obtained by minimizing with respect to $\vec{P}$.
However, in the experiment one uses an alternating ac-electric field and
measures the alternating portion of the charge on the capacitance plates.
So from the derivative of $\Phi'$ with respect to $\vec{P}$ we can disregard the terms that are
related to the ME$_H$ effect. This gives
\be
E^\alpha_{ac} = \frac{P_{ac}^\alpha}{\chi_e} - \zeta_0 \frac{\partial \mc{U}_0}{\partial P^\alpha}~.
\ee
With $E_{ac}^\alpha = \sum_\beta \chi^{-1}_{\alpha\beta} P_{ac}^\beta$, we get
\be
\delta\chi^{-1}_{\alpha\beta} \equiv  \chi^{-1}_{\alpha\beta}-\chi_e^{-1} \delta_{\alpha\beta} =
-\zeta_0 \frac{\partial^2 ~\mc{U}_0}{\partial P^\alpha \partial P^\beta}~.
\ee
Now, given that the magnetocapacitance is a very small quantity (MC $\ll$ 1\%), we may write
$\chi_e \bs{\chi}^{-1} \simeq{\bf 1} - \delta\bs{\chi}/\chi_e$, and thus $\delta\bs{\chi}/\chi_e \simeq - \chi_e\delta\bs{\chi}^{-1}$.
If we are measuring the charge $Q_s$ on a surface S (which is perpendicular to the applied ac-electric field), then
\be
\text{MC}\!=\!\frac{\delta\chi_{ss}}{\chi_e}
\!=\! \zeta_0 \chi_e \sum_{\alpha\beta} e_s^\alpha ~e_s^\beta  \frac{\partial^2 ~\mc{U}_0}{\partial P^\alpha \partial P^\beta}.
\ee

Taking derivatives we find
\be
\frac{\partial^2 \mc{U}_0}{\partial P^\alpha \partial P^\beta} = 2 \{\bs{\eta}_1,\bs{\eta}_1\}_+^\alpha~ \{\bs{\eta}_1,\bs{\eta}_1\}_+^\beta
\ee
For example, for $\alpha=\beta=x$ (which is directly relevant to most of the MC data we show here),
$\frac{\partial^2 \mc{U}_0}{\partial P_x^2 } = 2 \left( \eta_1^y \eta_1^z \right)^2$.
Taking the uniform portion of $\left( \eta_1^y \eta_1^z \right)^2$ for the conical helix gives
\be\label{eqn:mc1}
\frac{\text{MC}}{2 \chi_e \zeta_0} = f_1~(e_h^y e_h^z)^2 + f_2~(e_h^x)^2 + f_3~,
\ee
where
\bea
f_1(\vec{H}_a)&=& ( \eta_1^\parallel )^4 - 3 ( \eta_1^\parallel \eta_1^\perp )^2 + \frac{3}{8} (\eta_1^\perp)^4 \label{eqn:f1}\\
f_2(\vec{H}_a)&=&\frac{1}{8} (\eta_1^\perp)^4 - \frac{1}{2}(\eta_1^\parallel\eta_1^\perp)^2 \\
f_3(\vec{H}_a)&=& \frac{1}{2} (\eta_1^\parallel \eta_1^\perp)^2
\eea

\subsection{Angular dependence of MC: Comparison with exp. data}
We now consider the data shown in Fig.~\ref{MC100vsPhiHor}(b) for  the angular dependence of the MC at ``point B'',
when the field rotates in the (100) crystal plane.
Replacing $\vec{e}_h=\cos\phi [010]+\sin\phi[001]$ in Eq.~(\ref{eqn:mc1}) yields
\bea
\frac{\text{MC}}{2\chi_e \zeta_0} &=& \frac{1}{4} f_1 \sin^2(2\phi) + f_3~,
\eea
which is exactly the angular dependence observed in Fig.~\ref{MC100vsPhiHor}(b).
This provides confidence that the invariant $\mc{U}_0$ must be present in $\delta\Phi$.

Next, we consider the angular dependence shown in Fig.~\ref{MCvsAngle100katak}(b).
Replacing $\vec{e}_h=\cos\theta [100]+\sin\phi[03-7]/\sqrt{58}$ in Eq.~(\ref{eqn:mc1}) now gives
\be
\frac{\text{MC}}{2\chi_e \zeta_0} = \frac{21^2}{58} f_1 \sin^4\theta + f_2 \cos^2\theta + f_3 ~.
\ee
At the 1/2-plateau,  $f_1=1$, $f_2=f_3=0$ (since $\eta_1^\perp=0$), and so
this expression predicts that $\text{MC}\!\propto\!\sin^4\theta$, which is clearly in disagreement
with the $\sim \sin^2\theta$ behavior of the data.
So we may conclude that there is at least one more invariant giving a $\sim \sin^2\theta$ contribution.
In particular, this term must scale quadratically with the magnetic order parameter ($p=2$) to account for the angular dependence.
A list of such invariant terms is provided and discussed in App.~\ref{app:MC}.

\section{Summary and discussion}\label{sec:disc}
We have observed the ME effect in single crystals of the cubic compound Cu$_2$OSeO$_3$,
and we have studied its angular and temperature dependence.
We have demonstrated that the electric polarization can be controlled by an external magnetic field.
We have also investigated the temperature, field and angular dependence of the magnetocapacitance.
Both the magnetic field-induced polarization P and the magnetocapacitance MC set in at the magnetic ordering temperature
and show a number of anomalous features which we ascribe to magnetic phase transitions.
From the positions of these anomalies we have mapped out the phase diagram of the system in the $H_a$-T plane.
In the field region below the 3up-1down collinear ferrimagnetic phase and for $T<$30 K we find two additional magnetic phases, whose origin is
related to the presence of a Dzyalozinskii-Moriya chiral term in the free energy which impairs the homogeneity of
the magnetic ordering and gives rise to a continuous helical twisting of the ferrimagnetic order parameter with very long wavelength.
The vanishing of the ME response at the low-field phase of Cu$_2$OSeO$_3$ suggests a multi-domain structure of flat helices,
where the propagation vectors of the different helices are pinned along the preferred axes of the system.
In the intermediate-field phase the propagation vectors of the helices are aligned along the applied magnetic field $\vec{H}_a$.
For T$\gtrsim$ 30 K our magnetization measurements suggest a more complicated state and a possible thermal reorientation
of the anisotropy. Furthermore we also find a narrow range (2K)  close to the transition temperature $T_c$ with a double-peak anomaly in the ME$_H$
effect which might be due to the possible presence of the ``A-phase''.

We have also developed a Landau-Ginzburg phenomenological theory which accounts for the ME$_H$ effect in the intermediate-field and the 1/2-plateau phase,
and explains the angular, temperature and field dependence of this effect in a simple and physically transparent way.
Contrary to expectations, the influence of the DM coupling in the ME effect is heavily diminished due to the long-wavelength nature of the twist.
Instead, the ME effect is driven by an exchange striction mechanism involving the symmetric portion of the exchange anisotropy,
with three central features:
(i) the uniform component of $\vec{P}$ points along the vector $(H^yH^z, H^zH^x, H^xH^y)$;
(ii) its strength is proportional to $\eta_\parallel^2-\eta_\perp^2/2$, where $\eta_\parallel$ is the longitudinal and $\eta_\perp$
is the transverse (and spiraling) component of the magnetic ordering.
Hence, the field dependence of P provides a clear signature of the evolution of a conical helix under a magnetic field;
and (iii) apart from its uniform component, the polarization $\vec{P}$ has also a long-wavelength spatial component that is given by the pitch of the magnetic helices.
This effectively mesoscale antiferroelectric structure could in principle be probed by experiments with high spatial resolution,
which may further clarify the ME coupling mechanism in this Cu$_2$OSeO$_3$ at a more microscopic level.


\section{Acknowledgments}
We acknowledge experimental assistance from K. Schenk, and C. Amendola.
We would also like to thank A.N. Bogdanov, O. Janson, H. Rosner, K. Schenk, M. Wolf,
and S. Wurmehl for useful discussions.
One of the authors (H. B.) acknowledges financial support from the Swiss NSF and by the NCCR MaNEP.


\section*{appendix}
\subsection{Lifshitz invariants for the ME$_H$ effect}\label{app:ME}
Here we analyze the Lifshitz-type invariants between magnetic order parameters
and dielectric polarization.\cite{Baryakhtar,Stefanovskii,Mostovoy}
These (free) energy terms are linear in one spatial derivative of the magnetic order parameter
and are allowed in the present compound.
As we explained above and show also below, these ME terms are most likely not related to the observed ME$_H$ effect,
despite the fact that they give the same angular dependence with the term of Eq.~(\ref{eqn:fe}).
However, in a systematic expansion of the ME free energy they should not be omitted, as they yield terms linear in $\mathbf{P}$ and
may modify certain aspects of the ME behavior, in particular in systems with shorter pitch magnetic helices.

The Lifshitz invariants are the following seven terms
\bea
&& L_1 = \{\vec{P},\bs{\eta}_i\}_+ \cdot \{\bs{\partial},\bs{\eta}_j\}_+ , ~~ L_2 = \{\vec{P},\bs{\eta}_i\}_+ \cdot \{\bs{\partial},\bs{\eta}_j\}_- \nonumber \\
&& L_3 =\{\vec{P},\bs{\eta}_i\}_- \cdot \{\bs{\partial},\bs{\eta}_j\}_+ ,  ~~ L_4 =\{\vec{P},\bs{\eta}_i\}_- \cdot \{\bs{\partial},\bs{\eta}_j\}_- \nonumber \\
&& L_5 =P^x \eta_i^x \partial^x \eta_j^x + P^y \eta_i^y \partial^y \eta_j^y + P^z \eta_i^z \partial^z \eta_j^z \nonumber\\
&& L_6 =P^x \eta_i^x \partial^y \eta_j^y + P^z \eta_i^z \partial^x \eta_j^x + P^y \eta_i^y \partial^z \eta_j^z \nonumber\\
&& L_7 =P^x \eta_i^x \partial^z \eta_j^z + P^z \eta_i^z \partial^y \eta_j^y + P^y \eta_i^y \partial^x \eta_j^x \nonumber
\eea
and their contribution to the free energy is
\be
\delta\Phi_{L} = -\sum_{\alpha=1}^7 \lambda_\alpha L_\alpha~,
\ee
where $\lambda_1\!-\!\lambda_7$ are the corresponding coupling constants.
The contribution $\delta \vec{P}_{\!L}$
of these terms to the polarization can be found again by differentiating with respect to $\vec{P}$.
For example, for $\vec{E}=0$:
\bea
\frac{\delta P_{\!L}^x}{\chi_e} \!&=&\!  \lambda_1 \eta_i^y \partial^x \eta_j^y +  \lambda_2 \eta_i^y \partial^y \eta_j^x
+ \lambda_3 \eta_i^z \partial^z \eta_j^x + \lambda_4 \eta_i^z \partial^x \eta_j^z \nonumber \\
&+& \lambda_5 \eta_i^x \partial^x \eta_j^x + \lambda_6 \eta_i^x \partial^y \eta_j^y + \lambda_7 \eta_i^x \partial^z \eta_j^z ~.\nonumber
\eea
In order to account properly for the spatial derivatives, we may rewrite the position of a given Cu site as
$\vec{r}_i=\vec{r}+\bs{\rho}_i$, where $\bs{\rho}_i$ denotes the relative position of that site
with respect to the center of the corresponding unit cell $\vec{r}$.
Now, from each term $\eta_i^\alpha \partial^\gamma \eta_j^\beta$ there will be two types of contributions,
one proportional to $k^\gamma$ and another proportional to $k^\gamma (\vec{k}\cdot\delta\bs{\rho}_{ij})$.
The latter will be disregarded in the following since it is quadratic in $k$.
With this in mind, and keeping only the uniform component
\bea
( \eta_i^\alpha \partial^\gamma \eta_j^\beta )^{\text{uni}} &=&
\frac{1}{2} ( k ~\eta_i^\perp \eta_j^\perp )~e_h^\gamma~\sum_\kappa  \epsilon_{\alpha\beta\kappa}   e_h^\kappa \nonumber
\eea
one can easily show that only $L_2$ and $L_3$ contribute to the uniform portion of $\delta P_{\!L}^x$, and
the same happens for $\delta P_{\!L}^y$ and $\delta P_{\!L}^z$. The final expression for $\delta\vec{P}_{\!L}$ is
\be
\delta\vec{P}_{\!L} = \chi_e~(\lambda_3-\lambda_2)~k~\eta_i^\perp \eta_j^\perp~\{ \vec{e}_h,\vec{e}_h \}_+,
\ee
which has exactly the same angular dependence as in Eq.~(\ref{eqn:p2}) above.
The difference is that here the pre-factor scales with $(\eta^\perp)^2$ which
cannot account for the observed field dependence, and in particular the sign change of $\vec{P}$ between
the ``point A'' and the 1/2-plateau (see Sec.~\ref{sec:theoryME}).

\subsection{Magnetocapacitance}\label{app:MC}
Here we provide a list of invariant terms that are quadratic in P and thus may contribute to the MC, in addition to
the term $\mc{U}_0$ that we discussed in Sec.~\ref{sec:theoryMC}. These invariants are quadratic
(in contrast to $\mc{U}_0$ which is quartic) in the magnetic order parameters and as such, they
provide the $\sim \sin^2\theta$ angular dependence that is needed
for the full interpretation of the data shown in Fig.~\ref{MCvsAngle100katak}(b).

We restrict ourselves to spatially uniform invariants. These are the following five
\bea
\mc{U}_1&=&\{ \bs{\eta}_i,\bs{\eta}_j \}_+ \cdot \{ \vec{P},\vec{P} \}_+ \nonumber \\
\mc{U}_2&=&\{ \bs{\eta}_i,\bs{\eta}_j \}_- \cdot \{ \vec{P},\vec{P} \}_+ \nonumber\\
\mc{U}_3&=&P_x^2 \eta_i^x\eta_j^x + P_z^2 \eta_i^z\eta_j^z+P_y^2 \eta_i^y\eta_j^y  \nonumber\\
\mc{U}_4&=&P_x^2 \eta_i^y\eta_j^y + P_z^2 \eta_i^x\eta_j^x+P_y^2 \eta_i^z\eta_j^z  \nonumber\\
\mc{U}_5&=&P_x^2 \eta_i^z\eta_j^z + P_z^2 \eta_i^y\eta_j^y+P_y^2 \eta_i^x\eta_j^x \nonumber
\eea
Including these invariants in Eq.~(\ref{eqn:feMC1}) gives
\be\label{eqn:feMC2}
\delta\Phi = \Phi'-\Phi = -\zeta_0 ~\mc{U}_0  - \sum_{\alpha=1}^5 \zeta_\alpha ~\mc{U}_\alpha
\ee
where $\zeta_1\!-\!\zeta_6$ are the new coupling constants.
Repeating the steps described in Sec.~\ref{sec:theoryMC} leads to
\bea
\frac{\partial^2 \delta\Phi}{\partial P_x^2} \!=\! -2 \zeta_0 \left( \eta_1^y \eta_1^z \right)^2
-2\Big(  \zeta_3 \eta_i^x\eta_j^x + \zeta_4 \eta_i^y\eta_j^y + \zeta_5 \eta_i^z\eta_j^z \Big),\nonumber
\eea
where the second term gives the contributions from the invariants $\mc{U}_1$-$\mc{U}_5$. Using
\bea
( \eta_1^\alpha \eta_1^\beta )^{\text{uni}}&=&\Big[ (\eta_1^\parallel)^2 - \frac{1}{2} (\eta_1^\perp)^2 \Big] e_h^\alpha e_h^\beta + \delta_{\alpha\beta} \frac{1}{2} (\eta_1^\perp)^2, \nonumber
\eea
we find that the contribution to the MC from $\mc{U}_1$-$\mc{U}_5$ is
\bea
\frac{\text{MC}'}{2\chi_e\sigma_{ij}} &=&  \left[ \zeta_3 (e_h^x)^2 + \zeta_4 (e_h^y)^2 + \zeta_5 (e_h^z)^2 \right] \cdot
\left[ (\eta_1^\parallel)^2- \frac{1}{2}(\eta_1^\perp)^2 \right] \nonumber\\
&+& \frac{1}{2}(\zeta_3+\zeta_4+\zeta_5) (\eta_1^\perp)^2~.~~~~~
\eea
where $\sigma_{ij}=\pm 1$ is again the sign of $\eta_i\eta_j$.


\end{document}